\documentclass[tightenlines,11pt,titlepage]{revtex4}%
\pdfoutput=1
\usepackage{amsfonts}
\usepackage{amsmath}
\usepackage{amssymb}
\usepackage{graphicx}%
\begin{document}
\title{Optical M\"{o}bius Strips in Three Dimensional Ellipse Fields: Lines of Linear
Polarization }
\author{Isaac Freund}
\affiliation{Department of Physics, and Jack and Pearl Resnick Advanced Technology
Institute, Bar-Ilan University, Ramat-Gan 52900, Israel}

\begin{abstract}
The minor axes of, and the normals to, the polarization ellipses that surround
singular lines of linear polarization in three dimensional optical ellipse
fields are shown to be organized into M\"{o}bius strips and into structures we
call \textquotedblleft rippled rings\textquotedblright\ (r-rings). \ The
\ M\"{o}bius strips have two full twists, and can be either right- or
left-handed. \ The major axes of the surrounding ellipses generate cone-like
structures. Three orthogonal projections that give rise to $15$ indices are
used to characterize the different structures \ These indices, if independent,
could generate $839,808$ geometrically and topologically distinct lines;
selection rules are presented that reduce the number of lines to $8,248$, some
$5,562$ of which have been observed in a computer simulation. \ Statistical
probabilities are presented for the most important index combinations in
random fields. \ It is argued that it is presently feasible to perform
experimental measurements of the M\"{o}bius strips, r-rings, and cones
described here theoretically. \ \ \

\end{abstract}
\maketitle

\section{INTRODUCTION}

We describe here the M\"{o}bius strips, and other structures, generated by the
axes of the polarization ellipses that surround singular lines of linear
polarization (L lines) in three dimensional (3D) optical ellipse fields.
\ These strips have four half-twists; they complete the trinity of
two-half-twist Mobius strips that surround ordinary ellipses [$1,2$], and the
one-half- and three-half-twist strips that surround singular lines of circular
polarization (C lines) [$3$].

\begin{figure}
[h]
\includegraphics[width=0.99\textwidth]%
{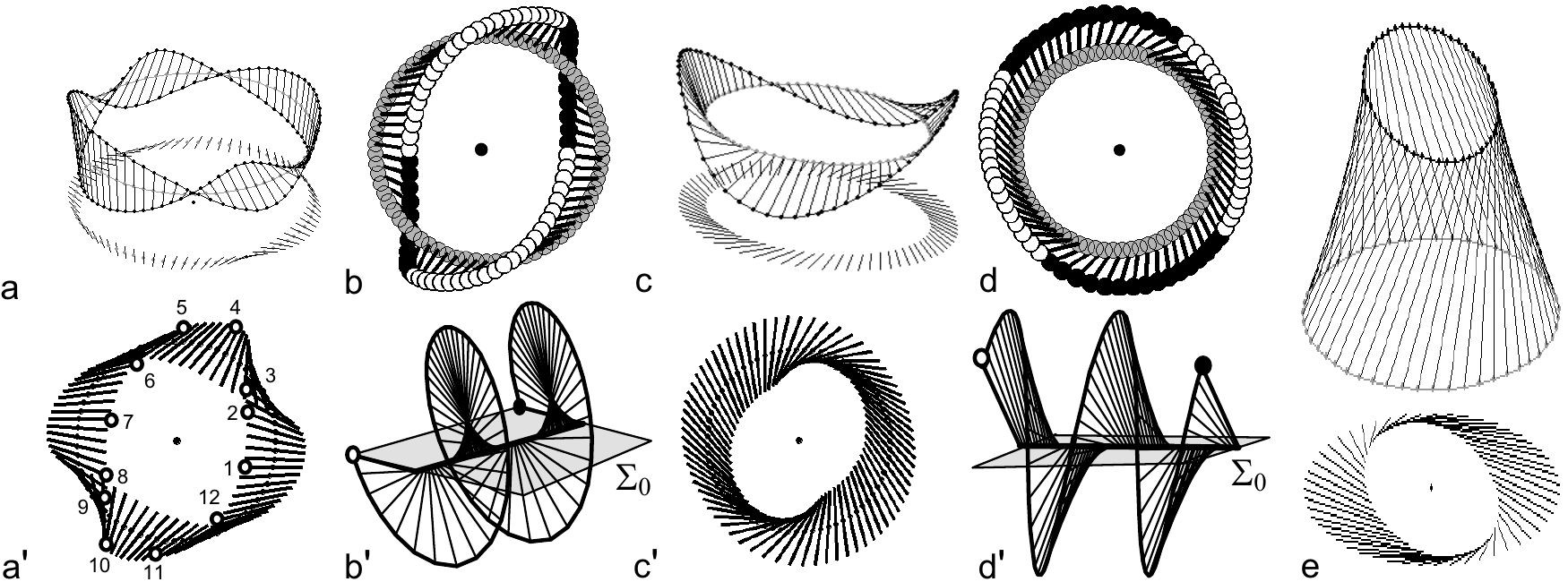}%

\caption{Typical structures surrounding L lines in a (here computed) random 3D
ellipse field. \ Shown in (a) $-$ (d') (in (e)) are the structures generated
by the minor (the major) axes of the ellipses whose centers are located in a
plane $\Sigma_{0}$ on a small circle that surrounds the line. (a) 3D view of a
M\"{o}bius strip that surrounds an L line shown floating above its projection
onto $\Sigma_{0}$. \ (a') Enlarged view of the projections of the ellipse axes
in (a) onto $\Sigma_{0}$. \ Along a counterclockwise path ($1-12$) the axis
projections rotate through $360^{o}$ in the retrograde, clockwise direction,
and $I=-1$. \ Other M\"{o}bius strips have $I=+1.$ \ (b) The strip in (a)
viewed from above. \ For clarity, here and in (b'), (c), (d), and (d') only
half of each axis is shown. \ Here and in (d) ellipse centers are shown by
gray circles and their minor axes by short straight lines; axis endpoints that
lie above (below) the plane of the circle of ellipse centers are shown by
filled white (black) circles. \ As can be seen, ellipse endpoints and centers
form a pair of interlocking rings with four crossings. \ \ (b') The closed
M\"{o}bius strip in (a),(b) opened to better display the double twist
structure. \ Here the ellipse centers form a straight line (the thick line)
around which the axes of the surrounding ellipses (thin lines) rotate. \ The
small white (black) circles mark the arbitrary first (last) ellipse. \ (c) 3D
view of a second type of structure that surrounds L lines, shown floating
above its projection onto $\Sigma_{0}$. \ This structure, which we call a
rippled ring (r-ring), is not a M\"{o}bius strip. \ (c') Enlarged view of the
axis projections onto $\Sigma_{0}$ of the r-ring in (c); the projections
rotate through $360^{o}$ in the same direction as the path, and $I=+1$.
\ Other r-rings have $I=-1$. \ (d) The r-ring in (c) viewed from above; the
endpoints and centers form noninterlocking rings. \ (d') The closed \ r-ring
in (c),(d) opened to display the ripple of the axes, which oscillate through
two complete sinusoidal cycles. \ In a random field approximately $1/3$ of all
L lines are surrounded by M\"{o}bius strips, $2/3$ by rippled rings. \ The
normals to the surrounding ellipses (not shown) also generate M\"{o}bius
strips and r-rings with $I=\pm1$. \ (e) The major axes of the ellipses in
$\Sigma_{0}$ generate cone-like structures, $\alpha$-cones, an example of
which is shown floating above its projection onto $\Sigma_{0}$; here $\ I=+1$,
other $\alpha$-cones have $I=-1$.}%
\label{Fig1}%
\end{figure}

On an L line the polarization ellipse collapses to a line, the major axis of
the ellipse, and the minor axis and ellipse normal become undefined
(singular). \ All three axes $-$ major, minor, and ellipse normal $-$ remain
well defined for the ellipses that \emph{surround} the L line. \ In a plane
$\Sigma$ pierced by an L line a point of linear polarization, an L point,
appears. \ The projections onto $\Sigma$ of the minor axes and ellipse normals
of the surrounding ellipses rotate about the point with winding number (net
rotation angle divided by $2\pi$) $I=\pm1$ [$4-13$]. \ Typical structures that
surround L lines are shown in Fig. \ref{Fig1}.%

Different types of L lines are characterized by the geometrical and
topological indices of the structures that surround the lines. \ Topological
index $I=\pm1$ generates two distinct types of L lines [$4,7,11-13$]; the
Mobius strips and other structures introduced here greatly expand the number
of these lines. \ In later sections we use $15$ different indices to
characterize these structures; these $15$ indices, if independent, could
generate $839,808$ different L lines. \ This number is drastically reduced to
$8,248$ by the selection rules described later; in a database containing
$10^{6}$ independent realizations of a simulated random ellipse field we find
some $5,562$ different L lines. \ The true number of different lines is likely
to be significantly greater, because as discussed later, due to the existence
of highly improbable configurations the yield of different lines grows so
slowly with the number of realizations that even $10^{9}$ independent
realizations may be insufficient to generate all allowed possibilities.

The plan of this report is as follows. \ In Section II we describe our
computer simulations and the numerical and analytical tools we use in later
sections to study L lines, their M\"{o}bius strips, and related structures.
\ In Section III we describe in detail the M\"{o}bius strips, r-rings, and
$\alpha$-cones (Fig. \ref{Fig1}) that surround L lines, and introduce $14$ new
indices to characterize these new structures. \ In Section IV we discuss the
selection rules that constrain the number of possible index combinations, and
present statistical probabilities for allowed combinations. \ We summarize our
findings in the concluding Section V. \ Throughout, as in [$1,3$], our
approach is descriptive, a more mathematical treatment similar to [$2$] will
be presented elsewhere.

\section{METHODS}

We briefly review here the methods we use to study L points on L lines. \ To a
large extent these methods are the same as those used in [$3$] to study C
points on C lines, and the reader may find it helpful to consult this
reference for more details. \ Many of the concepts used here and in [$3$] have
their origin in [$1,2$], and the reader may also find it helpful to consult
these references.

The principal axis system for the general polarization ellipse is the (here
right-handed) orthogonal three-frame $\boldsymbol{\alpha},\boldsymbol{\beta
},\boldsymbol{\gamma,}$ where $\boldsymbol{\alpha}$ and $\boldsymbol{\beta}$
are unit vectors directed along the major and minor ellipse axes, and
$\boldsymbol{\gamma}$ is a unit vector directed along the ellipse normal; the
positive end of $\boldsymbol{\gamma}$ can be uniquely defined by a right-hand
rule based on the rotation of the electric field vector $\mathbf{E}$ as it
traces out the ellipse over an optical cycle.

As noted above, at an L point on an L line there is only one principal axis,
$\boldsymbol{\alpha}$, that is well defined. \ Normal to this axis is a plane,
the principal plane here labelled $\Sigma_{0}$. \ In what follows, we consider
the 3D structure of the ellipses whose centers lie in $\Sigma_{0}$ on a small
circle $\sigma_{0}$ that surrounds the L point. \ We take the $z$-axis of the
fixed, orthogonal coordinate system to be along $\boldsymbol{\alpha}$, and the
corresponding $xy$-axes to lie in $\Sigma_{0}$, the orientation of these axes
in this plane being arbitrary.

As the plane of observation $\Sigma$ is tilted relative to $\Sigma_{0}$ the 3D
arrangement of the ellipses whose centers lie in $\Sigma$ changes. \ For very
small tilt angles these changes affect only the geometries of the structures,
but not their topologies, or their statistical properties. \ However, as the
rotation angle increases past some small, critical value that differs for each
L point, both the topology and the statistics change importantly. \ A similar
phenomenon occurs for points of circular polarization, C points, on C lines
[$3$]. \ All results presented in Fig. \ref{Fig1} (and unless stated otherwise
in all other figures) are for the case $\Sigma=\Sigma_{0}$. \ The complex set
of transformations that occur when $\Sigma$ is rotated away from $\Sigma_{0}$
will be reported on separately.

In general, $\boldsymbol{\alpha}$, and therefore $\Sigma_{0}$, make arbitrary
angles with the L line itself, so that in moving along the line from one L
point to another the coordinate system must be rotated in order to remain in
the principal plane $\Sigma_{0}$ of each L point.

\textbf{$\boldsymbol{\alpha}$}, \textbf{$\boldsymbol{\beta}$}, and
\textbf{$\boldsymbol{\gamma}$} can be calculated from the (here complex)
optical field $\mathbf{E}$ using either
\begin{subequations}
\label{alphabetagamma}%
\begin{align}
\boldsymbol{\alpha}  &  =\operatorname{Re}(\mathbf{E}^{\ast}\sqrt
{\mathbf{E}\boldsymbol{\cdot}\mathbf{E}}),\label{alpha}\\
\boldsymbol{\beta}  &  =\operatorname{Im}(\mathbf{E}^{\ast}\sqrt
{\mathbf{E}\boldsymbol{\cdot}\mathbf{E}}),\label{beta}\\
\boldsymbol{\gamma}  &  =\operatorname{Im}\left(  \mathbf{E}^{\ast
}\mathbf{\times E}\right)  , \label{gamma}%
\end{align}
\end{subequations}
which is due to Berry [$11,13$], or from the eigenvalues, $\lambda_{i}$, and
eigenvectors, $\boldsymbol{\nu}_{i}$, $i=1,2,3$, of the $3\times3$ real
coherency matrix [$14$] $M_{ij}=\operatorname{Re}(E_{i}^{\ast}E_{j}%
),\;i,j=x,y,z$. \ $\boldsymbol{\gamma}$ as defined in Eq. (\ref{gamma})
measures the area of the polarization ellipse and goes to zero at an L point.
\ At an L point the polarization is linear and $\mathbf{E}$ can be made be
pure real so that also $\boldsymbol{\beta}$ vanishes at the point. \ In what
follows, \textbf{$\boldsymbol{\alpha}$}, \textbf{$\boldsymbol{\beta}$}, and
\textbf{$\boldsymbol{\gamma}$ }for the surrounding ellipses are, without
change in notation, always normalized to unit length, i.e.
\textbf{$\boldsymbol{\alpha\Rightarrow\alpha}/\left\vert
\mathbf{\boldsymbol{\alpha}}\right\vert $}, etc.

We study two computed 3D ellipse fields. \ The first is composed of a large
number of linearly polarized plane waves with random propagation and
polarization directions, and random phases [$1$]. \ This field is an exact
solution of Maxwell's equations and serves as an important check on the
structures found using the more convenient linear expansion described below.
\ L lines in this field were traced out using the L point discriminant
$D_{L}=a_{2}$ obtained from the characteristic equation $\lambda^{2}%
+a_{1}\lambda+a_{2}=0$ of $M_{ij}$ [$15$].

In the immediate vicinity of an L point the field describing the ellipses in
$\Sigma_{0}$ can be expanded as

\begin{subequations}
\label{LpointE}%
\begin{align}
E_{x}  &  =\left(  P_{xx}+\text{i}Q_{xx}\right)  x+\left(  P_{xy}%
+\text{i}Q_{xy}\right)  y,\label{LpointE_a}\\
E_{y}  &  =\left(  P_{yx}+\text{i}Q_{yx}\right)  x+\left(  P_{yy}%
+\text{i}Q_{yy}\right)  y,\label{LpointE_b}\\
E_{z}  &  =a+\left(  P_{zx}+\text{i}Q_{zx}\right)  x+\left(  P_{zy}%
+\text{i}Q_{zy}\right)  y, \label{LpointE_c}%
\end{align}
\end{subequations}

where the direction of polarization of the point is along the $z$-axis and
$\Sigma_{0}$ is the $xy$-plane.

In many cases simpler expansions suffice: the M\"{o}bius strip in Fig.
\ref{Fig1}a,b\ is closely approximated by the field model $E_{x}=ix+\left(
1-i\right)  y,E_{y}=-iy,E_{z}=-1$, the r-ring in Fig. \ref{Fig1}c,d by
$E_{x}=-\left(  1+i\right)  x,E_{y}=ix+(1-i)y,E_{z}=1$, and the $\alpha$-cone
in Fig. \ref{Fig1}e by $E_{x}=-x,E_{y}=x-y,E_{z}=1$.

\begin{figure}
[h]
\includegraphics[width=0.75\textwidth]%
{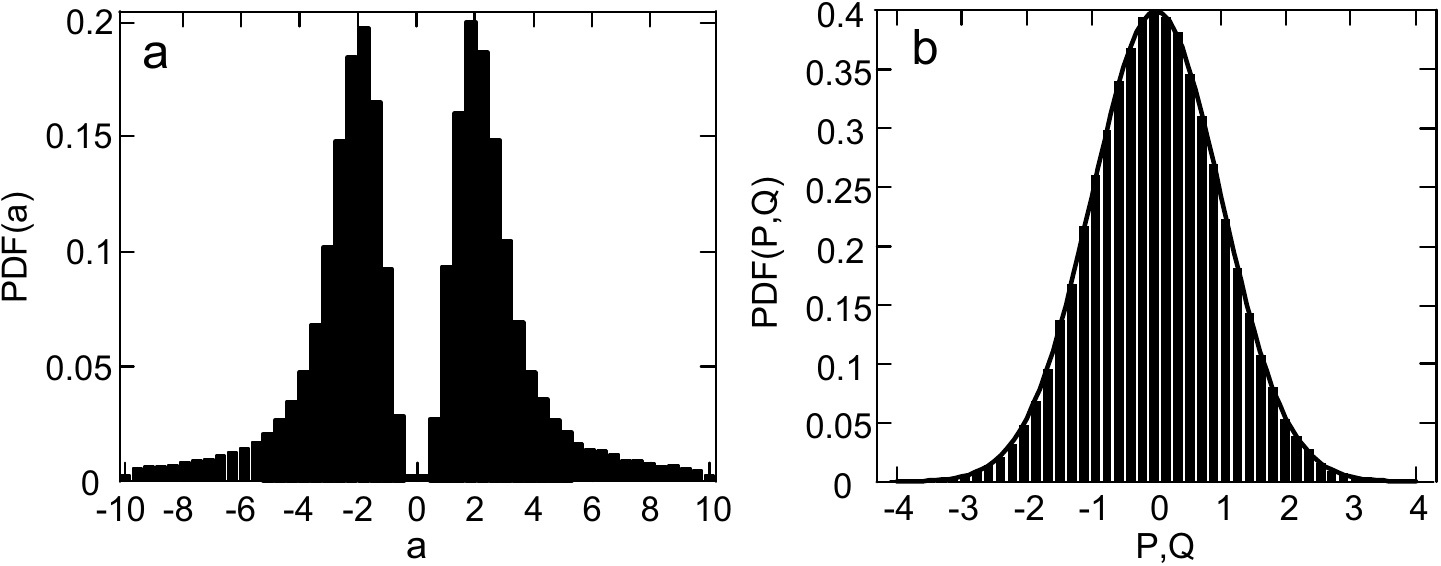}%
\caption{PDFs in $\Sigma_{0}$ of the parameters $a,P$, and $Q$, in Eq.
(\ref{LpointE}). \ (a) PDF of $a$. \ (b) PDF of $P,Q$. \ The numerical data
(histogram) closely matches a Gaussian with unit variance (curve). }%
\label{Fig2}%
\end{figure}

The statistics in $\Sigma_{0}$ of $a$ and the $P,Q$ in Eq. (\ref{LpointE})
were obtained using the numerical procedure described in [$3$] for C points,
modified for L points by the fact that whereas $\mathbf{E}\boldsymbol{\cdot
}\mathbf{E}=0$ for a C point, $\mathbf{E}^{\ast}\mathbf{\times E}=0$ for an L
point [$11-13$]. \ The probability density functions (PDFs) in $\Sigma_{0}$ of
$a$, and of $P,Q$ are shown in Fig. \ref{Fig2} $-$\ here all $P,Q$ have the
same PDF.%

\begin{figure}
[h]
\includegraphics[width=0.75\textwidth]%
{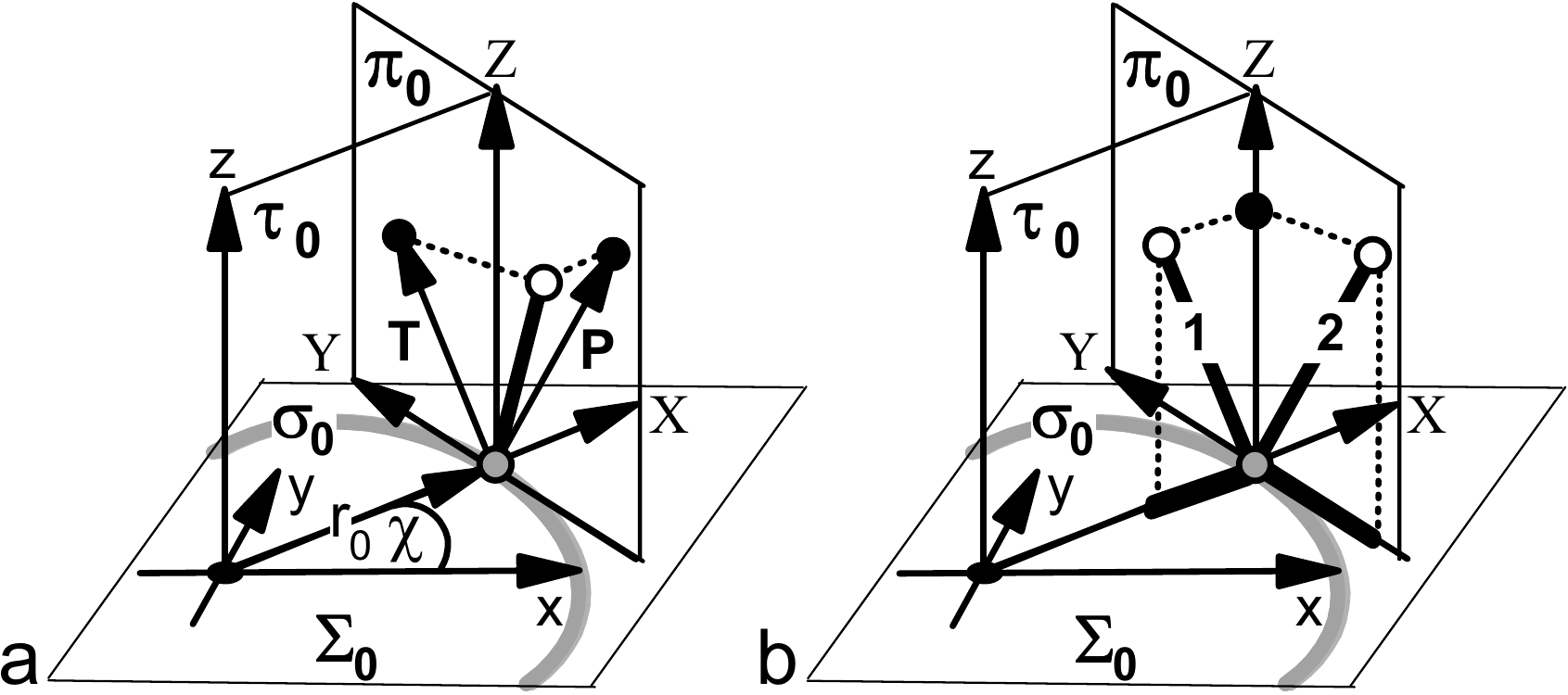}%
\caption{Projections $\tau_{0}$ and $\pi_{0}$. \ The L point is at the origin
of the fixed $xyz$-coordinate system. \ (a) Rotating plane $\tau_{0}$, the
$XZ$-plane ($\pi_{0}$, the $YZ$-plane), is radial (tangential) to $\sigma_{0}$
and travels on $\sigma_{0}$, rotating about the fixed $z$-axis; its position
on $\sigma_{0}$ is measured by the angle $\chi$, and by arc length
$s=r_{0}\chi$. \ An axis, $\boldsymbol{\alpha}$, $\boldsymbol{\beta}$, or
$\boldsymbol{\gamma}$, of the ellipse whose center (small gray circle) is
located at the origin of $XYZ$ is projected onto $\tau_{0}$ ($\pi_{0}$). \ The
endpoint of the axis is shown by a small white circle, the axis itself by a
thick line connecting the center and the endpoint. The endpoint projection in
$\tau_{0}$ (in $\pi_{0}$) is shown by a small black circle. \ The vector
\textbf{T} (\textbf{P}) in $\tau_{0}$ (in $\pi_{0}$) connects the ellipse
center with the endpoint projection. \ As $\tau_{0}$ ($\pi_{0}$) moves along
$\sigma_{0}$ different ellipses are projected onto $\tau_{0}$ ($\pi_{0}$),
their endpoints in $\tau_{0}$ ($\pi_{0}$) trace out a curve, the $\tau_{0}$
($\pi_{0}$) endpoint curve, and vector \textbf{T} (\textbf{P}) rotates. \ As
discussed in Section III, the winding numbers of rotation of \textbf{T}
(\textbf{P}) and of the tangent to the endpoint curve in $\tau_{0}$ (in
$\pi_{0}$), give rise to indices $\tau_{i}$ and $d\tau_{i}$ ($\pi_{i}$ and
$d\pi_{i}$)$,i=\alpha,\beta,\gamma$, that characterize the 3D arrangement of
the ellipses on $\sigma_{0}$. \ \ (b) Special projections. \ When the endpoint
projection onto $\tau_{0}$ (onto $\pi_{0}$) of an ellipse axis lies on the
$Z$-axis, as shown here for an axis of ellipse $1$ (ellipse $2$), the axis
projection onto $\Sigma_{0}$ is radial to $\sigma_{0}$, i.e. along the
$X$-axis (is tangential to $\sigma_{0}$, i.e. along the $Y$-axis). \ These two
important special cases are elaborated on in Section III$.$}%
\label{Fig3}%
\end{figure}

A single projection, the $\Sigma_{0}$ projection, cannot, of course, fully
describe the complex 3D arrangement of the ellipses surrounding an L line, and
in addition to projecting the ellipses on $\sigma_{0}$ onto $\Sigma_{0}$ we
study two additional projections $-$ $\tau_{0}$ and $\pi_{0}$. \ These
projections, which are the same as those used in [$1-3$], are reviewed in Fig.
\ref{Fig3}. \ Thus, in characterizing the 3D arrangement of the ellipses in
the vicinity of an L line we use three orthogonal projections, $\Sigma_{0}$,
$\tau_{0}$, and $\pi_{0}$ $-$ the minimum number required for a complex object.

\vspace*{0.2in}%

Issues regarding scaling and other technical aspects of the L point graphics
presented here are the same as those for the C point graphics presented in
[$3$]; the reader will find a full discussion of these issues in [$3$].

Unless stated otherwise, throughout we use a single value for the radius
$r_{0}=10^{-4}$ of the surrounding circle $\sigma_{0}$. \ This value is
sufficiently small that except for a scale factor the wavefield structure is
invariant under further reduction of the radius. \ The fact that every small
value for $r_{0}$ yields the same structure implies that the L point is
surrounded by \emph{nested} M\"{o}bius strips, r-rings, $\alpha$-cones, etc.
\ The full 3D arrangements of these deeply nested structures are, of course,
quite impossible to visualize; accordingly, below we dissect out a single,
typical surrounding circle, the $\sigma_{0}$ circle with $r_{0}=10^{-4}$, and
proceed to study its properties in detail.

\newpage

\section{INDICES OF M\"{O}BIUS STRIPS, r-RINGS, AND $\alpha$-CONES}

Here we discuss the $12$ winding numbers that describe the M\"{o}bius strips,
r-rings, and $\alpha$-cones that surround L lines, as well as the line
classification [$16$] and its relationship to these indices.

\vspace*{0.3in}

\subsection{Indices of the Projection onto $\Sigma_{0}$: $I_{\beta,\gamma}$
and $I_{\alpha}$}

\vspace*{0.1in}

As a result of the studies of Nye and coworkers [$4-10$], it has long been
known that the winding number of the projection of axis $\boldsymbol{\beta}$
onto $\Sigma_{0}$, $I_{\beta}=\pm1$ (Figs. \ref{Fig1}a',c'), equals
$I_{\gamma}$, the corresponding winding number of axis $\boldsymbol{\gamma}$.
\ The reason for the equivalence of these two indices is that as the radius of
the surrounding circle shrinks towards zero, by continuity the major axes of
the ellipses on this circle all approach parallelism with the direction of
polarization of the L point. \ The projections of $\boldsymbol{\beta}$ and
$\boldsymbol{\gamma}$ onto $\Sigma_{0}$ are therefore substantially
orthogonal, and as one rotates so does the other. \ Accordingly, in what
follows we use the symbol $I_{\beta,\gamma}$ to denote the common winding
number of these two axes. \ Although perhaps less obvious, the equivalence
$I_{\beta}=I_{\gamma}$ holds also for arbitrary $\Sigma$ [$13$].

The winding number $I_{\alpha}$ of the projection of axis $\boldsymbol{\alpha
}$ onto $\Sigma_{0}$ does not appear to have been discussed previously. \ We
find that just like for $I_{\beta,\gamma}$, in all cases $I_{\alpha}=\pm1$
(Fig. \ref{Fig1}e).

$I_{\alpha}$ and $I_{\beta,\gamma}$ are independent, and we find all four
possible sign combinations in our simulations.

It is not difficult to derive analytical expressions for the above indices.
\ Writing $\mathbf{V}$ for axis $\boldsymbol{\alpha}$, $\boldsymbol{\beta}$,
or $\boldsymbol{\gamma}$, we have from Eqs. (\ref{alphabetagamma}) and
(\ref{LpointE}) for the $xy$-components of $\mathbf{V}$ for sufficiently small
$x,y$,
%\end{subequations}

\begin{subequations}
\label{Vxy}%
\begin{align}
V_{x}  &  =F_{xx}x+F_{xy}y,\label{Vx}\\
V_{y}  &  =F_{yx}x+F_{yy}y, \label{Vy}%
\end{align}
\end{subequations}

where for

\begin{subequations}
\label{Fxxetc}%
\begin{align}
\text{axis }\boldsymbol{\alpha}\text{ }  &  \text{: \ }F=P\text{ \ (i.e.
}F_{xx}=P_{xx}\text{, etc.)};\label{alphaF}\\
\text{axis }\boldsymbol{\beta}\text{ }  &  \text{: \ }F=Q;\label{betaF}\\
\text{axis }\boldsymbol{\gamma}\text{ }  &  \text{: \ }F_{xx}=-2aQ_{yx}%
,F_{xy}=-2aQ_{yy},F_{yx}=2aQ_{xx},F_{yy}=2aQ_{xy}. \label{gammaF}%
\end{align}
\end{subequations}

Berry [$13$] has given the formula $I=$ sign$\left(  F_{xx}F_{yy}-F_{xy}%
F_{yx}\right)  $ for a general vector field of the form in Eq. (\ref{Vxy}).
\ Inserting Eq. (\ref{Fxxetc}) into Berry's formula, we have

\begin{subequations}
\label{Ialphabetagamma}%
\begin{align}
I_{\alpha}  &  =\text{sign}\left(  P_{xx}P_{yy}-P_{xy}P_{yx}\right)
,\label{Ialpha}\\
I_{\beta}  &  =I_{\gamma}=I_{\beta,\gamma}=\text{sign}\left(  Q_{xx}%
Q_{yy}-Q_{xy}Q_{yx}\right)  . \label{Ibetagamma}%
\end{align}
\end{subequations}

We find this result to be in full agreement with the indices obtained numerically.

\newpage

\subsection{Indices of the Projections onto $\tau_{0}$ and $\pi_{0}$}

\subsubsection{Projections onto $\tau_{0}$}

\paragraph{Indices $\tau_{i}$ and $d\tau_{i}$, $i=\beta,\gamma$}

\subparagraph{M\"{o}bius strips}

Both axes $\boldsymbol{\beta}$ and $\boldsymbol{\gamma}$ of the surrounding
ellipses can generate M\"{o}bius strips that are either right- or left-handed
screws with two full turns.

Index $\tau_{i}$ measures the number of turns and their handedness (sign), and
is $\tau_{i}=+2$ ($\tau_{i}=-2$) for a left-handed (right-handed) M\"{o}bius
strip. \ As can be seen, the M\"{o}bius strip in Fig. \ref{Fig1}a$-$b' is a
left handed screw with $\tau_{\beta}=+2$.

Although not an obvious geometrical or topological necessity, we find in
$\Sigma_{0}$ that if $\tau_{\beta}=+2$ ($\tau_{\beta}=-2$) then $\tau_{\gamma
}\neq-2$ ($\tau_{\gamma}\neq+2$), and vice versa. \ If both axes
$\boldsymbol{\beta}$ and $\boldsymbol{\gamma}$ generate Mobius strips (which
need not, and does not, always occur), then the above rule implies that both
strips must have the same handedness.

Index $d\tau$ measures the rotation of the tangent to the endpoint curve.
\ For M\"{o}bius strips, and \emph{only} for M\"{o}bius strips, we find that
in all cases $d\tau_{i}=\tau_{i}$, i.e. when $\tau_{i}=+2$ ($\tau_{i}=-2$)
$d\tau_{i}=+2$ ($d\tau_{i}=-2$). \ Because $\tau_{\beta}=\tau_{\gamma}$ when
both axes generate M\"{o}bius strips, $d\tau_{i}=\tau_{i}$ implies that for
such paired strips $d\tau_{\beta}=d\tau_{\gamma}$.

\begin{figure}
[h]
\includegraphics[width=0.65\textwidth]%
{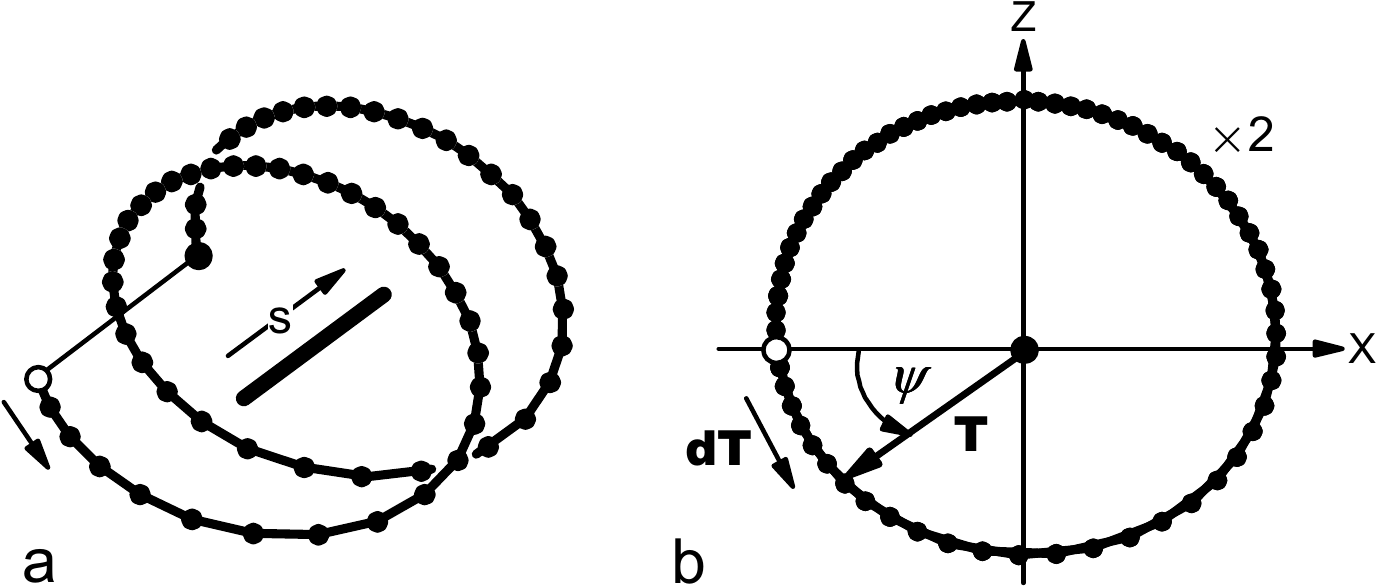}%
\caption{M\"{o}bius strip endpoint curves in $\tau_{0}$. \ Here and
throughout, small, connected black circles show the curve generated by the
projections of axis endpoints onto $\tau_{0}$\ as $\tau_{0}$ rotates on
$\sigma_{0}$ through one complete circuit around the L point; the starting
point of the circuit is shown by a small white circle; the sequence in which
the endpoint curve is drawn is indicated by a small arrow.\ (a),(b) $\tau_{0}$
endpoint curve of the M\"{o}bius strip in Figs. \ref{Fig1}a$-$b'. \ (a) 3D
view of the endpoint curve. \ The first and last points are shown connected
because they are the same point. \ The central straight line is formed by the
centers of the ellipses on $\sigma_{0}$. \ As can be seen, the endpoints
generate a two-turn left-handed helix. \ (b) The endpoint curve in (a) viewed
normal to $\tau_{0}$. The \textquotedblleft$\times2$\textquotedblright\ label
indicates that the second turn of the helix in (a) lies behind the first turn
visible here. \ Vector $\mathbf{T}$ has its origin at the center of the
ellipse on the surrounding circle $\sigma_{0}$ (the origin of $XZ$) and its
head at the endpoint projection. \ The winding number $\tau_{i}$, where for
the M\"{o}bius strip shown here $i=\beta$, is obtained from the rotation of
$\mathbf{T}$ by measuring the rotation angle $\psi\left(  s\right)
=\arctan\left(  T_{Z}\left(  s\right)  ,T_{X}\left(  s\right)  \right)  $, and
unwrapping (unfolding) the result to obtain $\tau_{i}=\left[  \psi\left(
2\pi\right)  -\psi\left(  0\right)  \right]  /\left(  2\pi\right)  $. \ The
winding number $d\tau_{i}$ of the tangent vector to the endpoint curve,
$\mathbf{dT}$, is similarly calculated from winding angle $\arctan\left(
dT_{Z}/ds,dT_{X}/ds\right)  $. \ Here the net total rotation of $\mathbf{T}$
and of $\mathbf{dT}$ is $+4\pi$, and $\tau_{i}=d\tau_{i}=+2$. \ Most endpoint
curves do not have a symmetrical, near circular shape, as does the one shown
here, which was chosen for simplicity and clarity of presentation. \ Although
here $\tau_{i}=d\tau_{i}$, in general these two indices can be
different.\ \ \ }%
\label{Fig4}%
\end{figure}

Indices $\tau_{i}$ and $d\tau_{i}$ for L point M\"{o}bius strips are
illustrated in Fig. \ref{Fig4}.%

\newpage

An important property of L point M\"{o}bius strips is that the first and
second turn of the two-turn helix are identical, so that the projections onto
$\tau_{0}$ of the two turns always overlap and only one turn is seen in such a
projection, Fig. \ref{Fig4}. \ We therefore include in such projections a
\textquotedblleft$\times2$\textquotedblright\ label to alert the reader to the
fact that there is a second turn underlying the first.

\vspace{-0.1in}

\subparagraph{r-rings}

\begin{figure}
[h]
\includegraphics[width=0.65\textwidth]%
{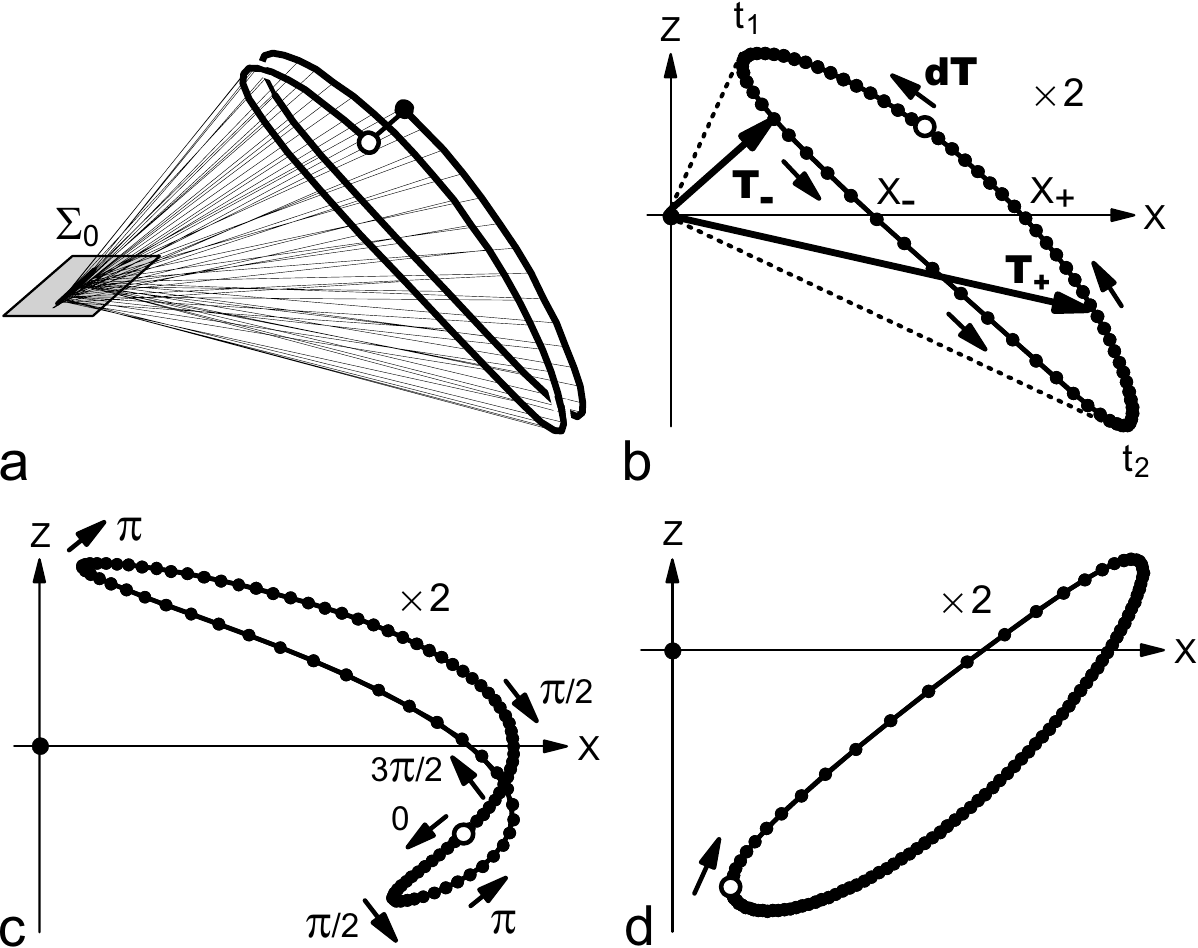}%
\caption{r-ring endpoint curves in $\tau_{0}$. \ (a) The opened r-ring in Fig.
\ref{Fig1}d' viewed from the left to display the two-turn helix generated by
the axis endpoints. \ (b) Endpoint curve of (a). \ Because the endpoints of an
r-ring do not spiral around the line of ellipse centers, the origin of the
vector $\mathbf{T}$ always lies outside the closed endpoint curve, and as a
result $\tau_{i}\equiv0$. \ This can be seen by following the rotation of
$\mathbf{T}$ or by using the phase ratchet rules. \ At \ turning points
$t_{1}$ and $t_{2}$ vector $\mathbf{T}$ reverses its direction of rotation,
which is counterclockwise, i.e. positive, (clockwise, i.e. negative) on the
right (left) half of the endpoint curve, and the net rotation of $\mathbf{T}$
is zero. \ In applying the phase ratchet rules [$3$], one first lists the
signed crossing of the endpoint curve with the $XZ$-axes in the order that
they occur, starting at the small white circle that marks the beginning of the
circuit on $\sigma_{0}$ and proceeding in the direction of the small arrow.
\ For the endpoint curve of the M\"{o}bius strip in Fig. \ref{Fig4}b the
crossing sequence is $Z_{+}X_{+}Z_{+}X_{+}$, where the sign of a crossing is
positive (negative) if $\mathbf{T}$ rotates counterclockwise (clockwise) at
the crossing. \ This is the canonical sequence for $\tau_{i}=+1$; multiplying
by $2$ to account for the hidden second turn yields the net index $\tau
_{i}=+2$. \ For the r-ring shown here the crossing sequence is $X_{-}X_{+}$.
\ Adjacent terms in the same axis necessarily have opposite signs and cancel,
so\ for this sequence $\tau_{i}=0$. \ The tangent vector, $\mathbf{dT}$,
however, is easily seen to rotate through $2\pi$ in the positive, clockwise
direction; multiplying by $2$, the net winding angle is $4\pi$ and $d\tau
_{i}=+2$. \ (c) An r-ring whose endpoint curve forms a figure of eight. \ Here
the crossing sequence is $X_{+}X_{-}$ and once again $\tau_{i}=0$, as it is
for every r-ring. \ But here also $d\tau_{i}=0$, as it is for every figure of
eight. \ That the net rotation of $\mathbf{dT}$ is zero can be seen by
following the tangent vectors marked with the cumulative rotation angle. \ (d)
An r-ring with $\tau_{i}=0$ and $d\tau_{i}=-2$. \ In a random field there are
roughly equal numbers of each type of r-ring: $\tau_{i}=0$ and $d\tau
_{i}=-2,0,+2$.\ }%
\label{Fig5}%
\end{figure}

The endpoint curves of r-rings form two-turn helices $-$ possibly surprising
because such a structure cannot be easily inferred from Figs. \ref{Fig1}%
c$-$d'. \ For r-rings, $\tau_{i}\equiv0$, whereas $d\tau_{i}=0,\pm2$.
\ Typical examples are shown in Fig. \ref{Fig5}, which also illustrates the
\textquotedblleft phase ratchet rules\textquotedblright\ that simplify
calculation of $\tau_{i}$ for complicated endpoint curves [$3$]. \ For paired
$\boldsymbol{\beta}-\boldsymbol{\gamma}$ r-rings $d\tau_{\beta}$ and
$d\tau_{\gamma}$ need no longer be equal, and we find in our simulations all
nine possible combinations of these two indices.%

\paragraph{Index $d\tau_{\alpha}$}

\begin{figure}
[h]
\includegraphics[width=0.85\textwidth]%
{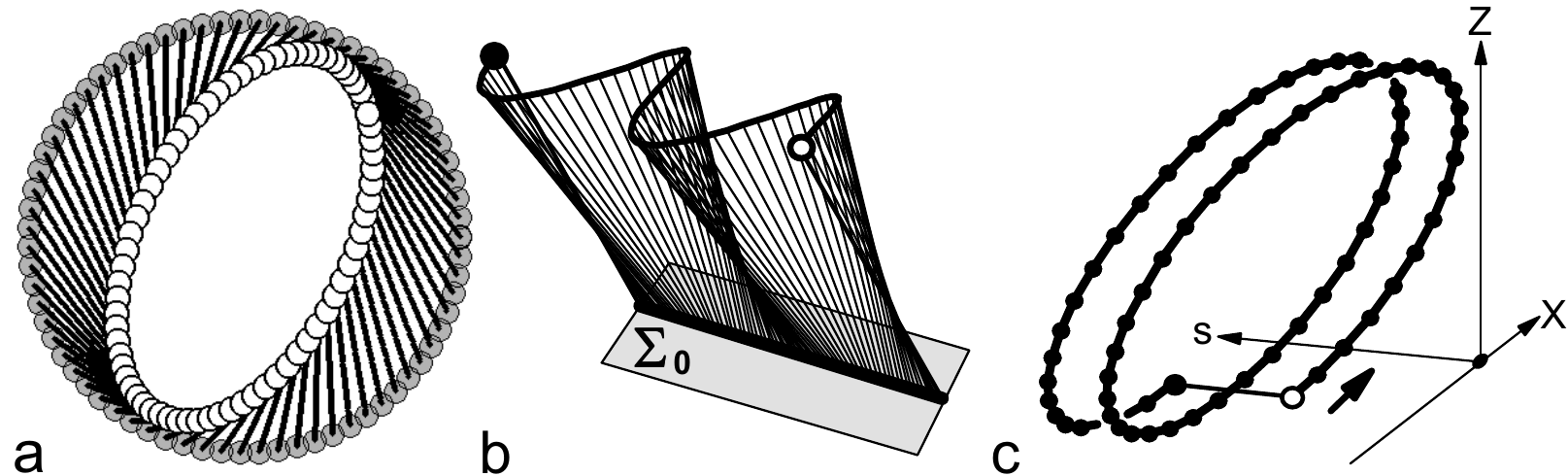}%
\caption{$\alpha$-cones. \ Shown are three additional views of the cone in
Fig. \ref{Fig1}e. \ (a) The cone viewed from above. \ (b) The cone in (a)
opened and straightened to show the oscillations of the axis endpoints. \ (c)
The $\tau_{0}$ projection of the axis endpoints. \ Here the $Z$-axis is
elongated relative to the $X$-axis to show the structure of the normally very
flat two-turn endpoint helix that is characteristic of $\alpha$-cones.}%
\label{Fig6}%
\end{figure}

As illustrated in Fig. \ref{Fig1}e, axes $\boldsymbol{\alpha}\ $of the
surrounding ellipses are not organized into M\"{o}bius strips or r-rings, but
rather into cone-like structures $-$ $\alpha$-cones. \ In Fig. \ref{Fig6} we
show additional views of the cone in Fig. \ref{Fig1}e. \ As can be seen, the
endpoint curve of an $\alpha$-cone is a two-turn helix, much like the helix of
an r-ring, and like an r-ring, $\tau_{\alpha}\equiv0$ also for the cone; once
again the reason is that the endpoint curve does not encircle the ellipse
centers and vector $\mathbf{T}$ therefore oscillates clockwise and
counterclockwise with a net winding angle of zero. \ Unlike r-rings, however,
the endpoint curves of $\alpha$-cones are never figures of eight, and for the
cones $d\tau_{\alpha}=\pm2$.

For the $\alpha$-cone in Fig. \ref{Fig6}a the axis endpoints form an ellipse
that is smaller than the circle of ellipse centers (i.e. the surrounding
circle). \ But for other cones, or for this cone viewed from below rather than
from above, the endpoint ellipse is larger than the circle. \ In either case
the two curves $-$ circle and ellipse $-$ never cross. \ There are, however,
numerous examples of cones where the major axis of the endpoint ellipse is
larger than, and the minor axis is smaller than, the diameter of the circle of
ellipse centers, and in these cases the two curves $-$ circle and ellipse $-$
cross four times.%

\subsubsection{Projections onto $\pi_{0}$}

\paragraph{Indices $\pi_{i}$ and $d\pi_{i}$, $i=\beta,\gamma$}

Indices $\pi_{i}$ and $d\pi_{i}$ are the $\pi_{0}$ projection analogs of
indices $\tau_{i}$ and $d\tau_{i}$ of the $\tau_{0}$ projection.

Index $\pi_{i}$ measures the rotation of the endpoint curve in $\pi_{0}$,
whereas index $d\pi_{i}$ measures the rotation of the tangent to this curve.
\ Fig. \ref{Fig4}\ is applicable to $\pi_{i}$ and $d\pi_{i}$ upon replacement
of the $X$-axis by the $Y$-axis, and vectors $\mathbf{T}$ and $\mathbf{dT}$ by
vectors $\mathbf{P}$ and $\mathbf{dP}$. $\ \pi_{i}$ and $d\pi_{i}$ can each
take on the values $0,\pm2$.

$\left\vert \tau_{\beta}\right\vert =2$ ($\left\vert \tau_{\gamma}\right\vert
=2$) implies that axis $\boldsymbol{\beta}$ (axis $\boldsymbol{\gamma}$)
generates a Mobius strip, because for r-rings (and cones) $\tau_{i}\equiv0$.
\ But the same is not true for $\pi_{i}$ which equals $\pm2$ if the endpoint
curve in $\pi_{0}$ encloses the ellipse centers, and is zero otherwise. \ In
very nearly $1/4$ of all r-rings $\left\vert \pi_{i}\right\vert =2$.

There exists a $\pi_{i}$ selection rule that is the analog of the $\tau_{i}$
selection rule noted above: when $\pi_{\beta}=+2$ ($\pi_{\beta}=-2$),
$\pi_{\gamma}\neq-2$ ($\pi_{\gamma}\neq+2$).

$d\pi_{i}$ is $\pm2$ for simple (non-self-intersecting curves), and is zero
for figures of eight. \ When $\pi_{i}=+2$ ($\pi_{i}=-2$), $d\pi_{i}=+2$
($d\pi_{i}=-2$).

\begin{figure}
[h]
\includegraphics[width=0.65\textwidth]%
{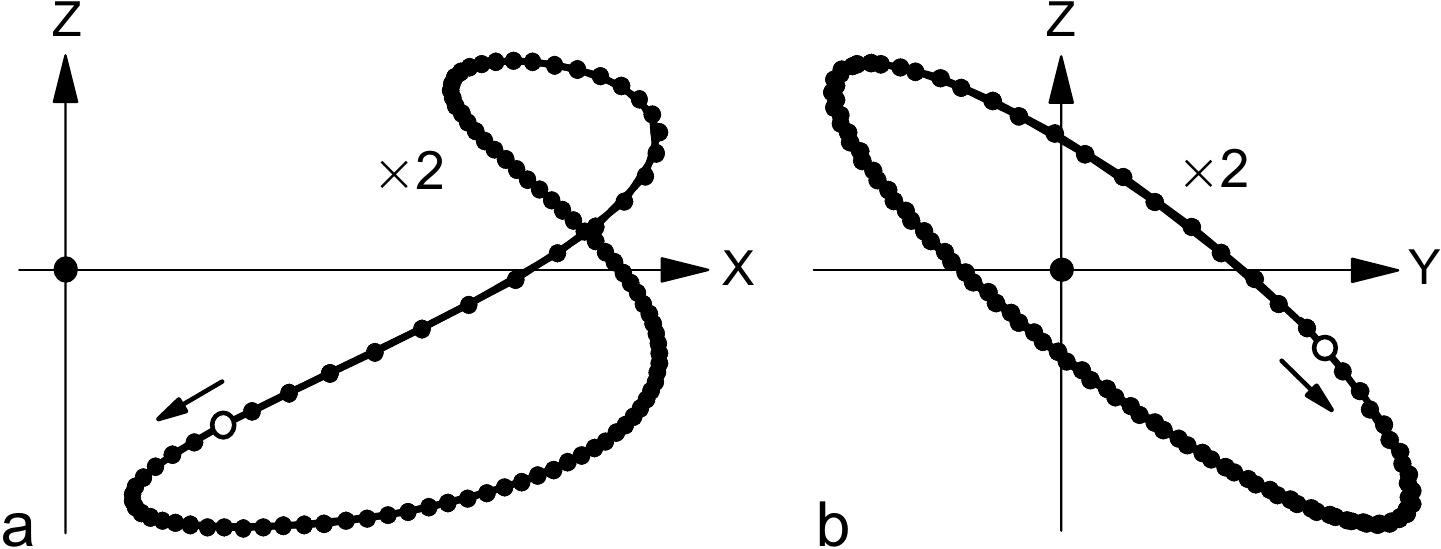}%
\caption{r-ring endpoint curves of axis $\boldsymbol{\gamma}$ in $\tau_{0}$
and in $\pi_{0}$. \ (a) $\tau_{0}$ endpoint curve. \ Both $\tau_{\gamma}$ and
$d\tau_{\gamma}$ are zero, as is always the case for a figure of eight. \ (b)
$\pi_{0}$ endpoint curve. \ The curve encloses the origin (circle of ellipse
centers), and both $\pi_{\gamma}$ and $d\pi_{\gamma}$ equal $-2$.}%
\label{Fig7}%
\end{figure}

Fig. \ref{Fig7} illustrates the case of an r-ring with $\tau_{\beta}%
=d\tau_{\beta}=0$, and $\pi_{\gamma}=d\pi_{\gamma}=-2$.

\bigskip

\paragraph{Index $d\pi_{\alpha}$}

Indices $\pi_{\alpha}$ and $d\pi_{\alpha}$ are the $\pi_{0}$ projection
analogs of indices $\tau_{\alpha}$ and $d\tau_{\alpha}$ of the $\tau_{0}$
projection. \ Like the endpoint curve of axis $\boldsymbol{\alpha}$ in
$\tau_{0}$, for which $\tau_{\alpha}\equiv0$ and $d\tau_{\alpha}=\pm2$, the
endpoint curve in $\pi_{0}$ generates a two-turn helix for which $\pi_{\alpha
}\equiv0$ and $d\pi_{\alpha}=\pm2$. \ Fig. \ref{Fig6}c for the $\tau_{0}$
projection is applicable also to the $\pi_{0}$ projection after the $X$-axis
is replaced by the $Y$-axis.

This completes our discussion of the $12$ indices that characterize the three
orthogonal projections, $\Sigma_{0}$, $\tau_{0}$, and $\pi_{0}$, of the Mobius
strips, r-rings and $\alpha$-cones that surround L lines. \ In the next
section we return to the $\Sigma_{0}$ projection and consider a
characterization [$16$] that is well known for C lines [$4,6,11-13$], but that
does not appear to have been discussed previously for L lines.

\vspace{0.2in}

\subsection{Line Classification: Indices $\Lambda_{i},i=\alpha,\beta,\gamma$}

\subsubsection{Computer Simulation}

The arrangement of the axis projections onto $\Sigma_{0}$ can be characterized
by two indices. \ The first, winding numbers $I_{\alpha}$ and $I_{\beta
}=I_{\gamma}=$ $I_{\beta,\gamma}$, measures the net rotation of the axis
projections around the L point, and has been discussed in Section III.A and
illustrated in Fig. \ref{Fig1}. \ The second, the line classification index
$\Lambda_{i},i=\alpha,\beta,\gamma$, counts the number of streamlines formed
by axis $i$ that terminate (or originate) on the L point as straight lines
(separatixes). \ For a sufficiently small surrounding circle $\sigma_{0}$, as
is used here, $\Lambda_{i}$ counts the number of axis projections on
$\sigma_{0}$ that point directly at the L point. \ We find that in all cases,
for positive (negative) $I_{i}$, $\Lambda_{i}=0,4$ ($4$). \ We note that
although the line classification [$16$] has not been previously discussed
explicitly for L points for any axis, these values are consistent with
streamline maps for axes $\boldsymbol{\beta}$ and $\boldsymbol{\gamma}$
presented in [$4,7,11$]; streamlines for axis $\boldsymbol{\alpha}$ do not
appear to have been considered previously.

\begin{figure}
[ph]
\includegraphics[width=0.8\textwidth]%
{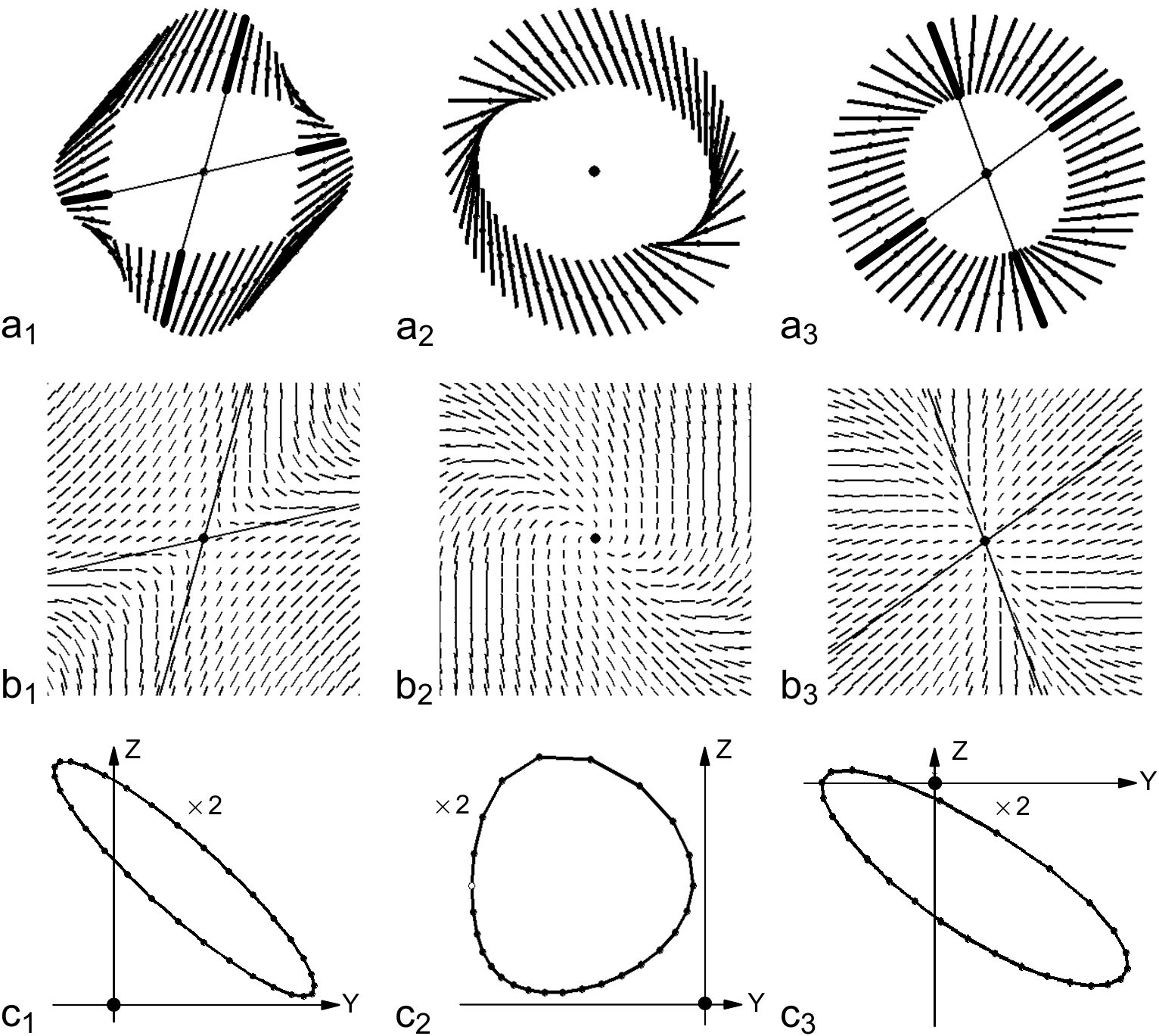}%
\caption{Line classification in a computed 3D random ellipse field. \ (a$_{1}%
$)$-$(a$_{3}$) Axis projections onto $\Sigma_{0}$. \ Shown as short straight
lines are the projections of the axes of the ellipses on the surrounding
circle $\sigma_{0}$. \ Projections that point directly at the circle center (L
point), here $\Lambda$-lines, are marked by thick lines; the line
classification index $\Lambda$ counts the number of these lines. \ In our
simulation $\Lambda$-lines are found numerically [$3$] as the solutions of
$\theta\left(  \chi\right)  =\chi+n\pi,$ with $n=-4...4$, as required. \ Here
$\theta$ is the angle that the axis projection makes with the $x$-axis, and
$\chi$ measures the position of the ellipse center on $\sigma_{0}$, Fig.
\ref{Fig3}. \ The straight lines passing through the L point are calculated
from Eq. (\ref{Yplusminus}). \ As can be seen, in each case shown here, and in
numerous other cases tested but not shown, there is complete agreement between
theory and the simulated field. (a$_{1}$) Axis $\boldsymbol{\alpha}$. \ Here
index $I_{\alpha}=-1$, $\Lambda_{\alpha}=4$; for other $\alpha$-cones
$I_{\alpha}=+1$, $\Lambda_{\alpha}=0,4$. \ (a$_{2}$) Axis $\boldsymbol{\beta}%
$. \ Here $I_{\beta}=+1$, $\Lambda_{\beta}=0$;\ for other $\boldsymbol{\beta}$
M\"{o}bius strips or r-rings $I_{\beta}=-1$,$+1$, $\Lambda_{\beta}=4$.
\ (a$_{3}$) Axis $\boldsymbol{\gamma}$. \ Here \ $I_{\gamma}=+1$,
$\Lambda_{\gamma}=4$;\ for other $\boldsymbol{\gamma}\ $M\"{o}bius strips or
r-rings $I_{\gamma}=-1$, $\Lambda_{\gamma}=4$, or $I_{\gamma}=+1$,
$\Lambda_{\gamma}=0$. \ (b$_{1}$)$-$(b$_{3}$) \ Axis projections onto
$\Sigma_{0}$ corresponding to the ellipse fields in (a$_{1}$)$-$(a$_{3}$),
respectively. \ Shown are axis projections for ellipses whose centers lie on a
square grid that surrounds the L point. \ These projections (short lines) are
the tangents to the streamlines; visually, their overall pattern approximates
the streamline pattern. \ The long straight lines are the separatixes
calculated from Eq. (\ref{Yplusminus}), and as can be seen are in full accord
with the patterns shown here, as well as with numerous other examples of these
patterns that are not shown. \ (b$_{1}$) This streamline pattern is similar to
the contour pattern surrounding a saddle point. \ \ (b$_{2}$) These
streamlines spiral around the L point. \ \ (b$_{3}$) Here the streamlines form
right and left leaning parabolic-like curves \textquotedblleft
centered\textquotedblright\ on the right leaning separatix; the fields of
right and left leaning \textquotedblleft parabolas\textquotedblright\ are
separated by the left leaning separatix. \ These three patterns are the only
ones found in our simulation. \ We find that each pattern is tied to a
particular combination of $I$ and $\Lambda$: in all of the numerous cases
tested for axes $\boldsymbol{\alpha}$, $\boldsymbol{\beta}$, and
$\boldsymbol{\gamma}$, when $I=-1$, $\Lambda=4$, the pattern is qualitatively
similar to that in (b$_{1}$); when $I=+1$ and $\Lambda=0$ the pattern is
qualitatively similar to that in (b$_{2}$); and when $I=+1$ and $\Lambda=4$
the pattern is qualitatively similar to that in (b$_{3}$). \ (c$_{1}$%
)$-$(c$_{3}$) Endpoint curves in $\pi_{0}$ corresponding to the axis
projections in (a$_{1}$)$-$(a$_{3}$). \ As can be seen, and as is expected
(Fig. \ref{Fig3}), $\Lambda$ equals the number of Z-axis crossings of the
endpoint curve in $\pi_{0}$. \ This has been verified for all $10^{6}$
realizations in our random field database. }%
\label{Fig8}%
\end{figure}

Unlike the case of winding number $I$ which is always the same for both axes
$\boldsymbol{\beta}$ and $\boldsymbol{\gamma}$, $\Lambda_{\beta}$ and
$\Lambda_{\gamma}$ can differ when $I_{\beta,\gamma}=+1$. \ Examples of the
line classification are illustrated in Fig. \ref{Fig8}. \ An important,
general property of these maps is that they posses a center of inversion.
\ This follows from the forms of Eqs. (\ref{alphabetagamma}) and
(\ref{LpointE}) for sufficiently small $x,y$. \ But this is the \emph{only}
intrinsic symmetry of such maps, and therefore of the corresponding streamlines.

As indicated in Fig. \ref{Fig3}, an axis projection points towards the L point
(circle center) only when the axis itself lies in the $XZ$-plane, so that the
projection of its endpoint onto $\pi_{0}$ lies on the $Z$-axis. \ Also this
property is illustrated in Fig. \ref{Fig8}. \ Thus, $\Lambda$ also equals the
number of times the endpoint curve in $\pi_{0}$ crosses the $Z$-axis, and vice versa.%

\newpage

\subsubsection{Theory}

The above results can be obtained analytically from the properties of the
streamlines. \ These streamlines, expressible as the family of curves
$y\left(  x\right)  $, are the solutions of%

\begin{equation}
\frac{dy}{dx}=\frac{V_{y}}{V_{x}}=\frac{F_{yx}x+F_{yy}y}{F_{xx}x+F_{xy}y},
\label{dydx}%
\end{equation}

where in writing the second form we have used Eq. (\ref{Vxy}); $F_{xx},F_{xy}%
$, etc., are given in Eq. (\ref{Fxxetc}).

Eq. (\ref{dydx}) is a classic exercise in the theory of ordinary differential
equations, and the properties of its solutions are well known [see for example
$17$]. \ There are six types of solution, three of which are unstable under
small perturbation, and are therefore nongeneric and do not appear in our
simulations, and three which are stable and which do appear. \ For
completeness, and in order to help the reader avoid a possible
misinterpretation of previous illustrations of $\boldsymbol{\beta
},\boldsymbol{\gamma}$ streamlines [$4,7,11$], below we consider briefly all
six types.

The solutions of Eq. (\ref{dydx}) are characterized by both index $I$ and the
discriminant%

\begin{equation}
\Delta=\left(  F_{xx}-F_{yy}\right)  ^{2}+4F_{xy}F_{yx}. \label{Delta}%
\end{equation}

When $\Delta<0$, $\Lambda=0$, when $\Delta=0$, $\Lambda=2$, and when
$\Delta>0$, $\Lambda=4$. \ As discussed below, the case $\Delta=0$,
$\Lambda=2$ corresponds to one of the unstable solutions and is not seen in
our simulation; the other two possibilities are.

The rule discussed above, that for $I_{i}=-1$, $\Lambda_{i}\equiv4$,
$i=\alpha,\beta,\gamma$, can be easily proven by rewriting Eq. (\ref{Delta})
as%

\begin{align}
\Delta &  =\left(  F_{xx}+F_{yy}\right)  ^{2}-4\mathcal{I},\nonumber\\
\mathcal{I}  &  =F_{xx}F_{yy}-F_{xy}F_{yx}. \label{negativeI}%
\end{align}

Recalling from Eq. (\ref{Ialphabetagamma}) that $I=$ sign$\left(
\mathcal{I}\right)  $, it immediately follows that if $I<0$, $\Delta>0$, and
as a result, $\Lambda\equiv4$.

\begin{figure}
[ph]
\includegraphics[width=0.90\textwidth]%
{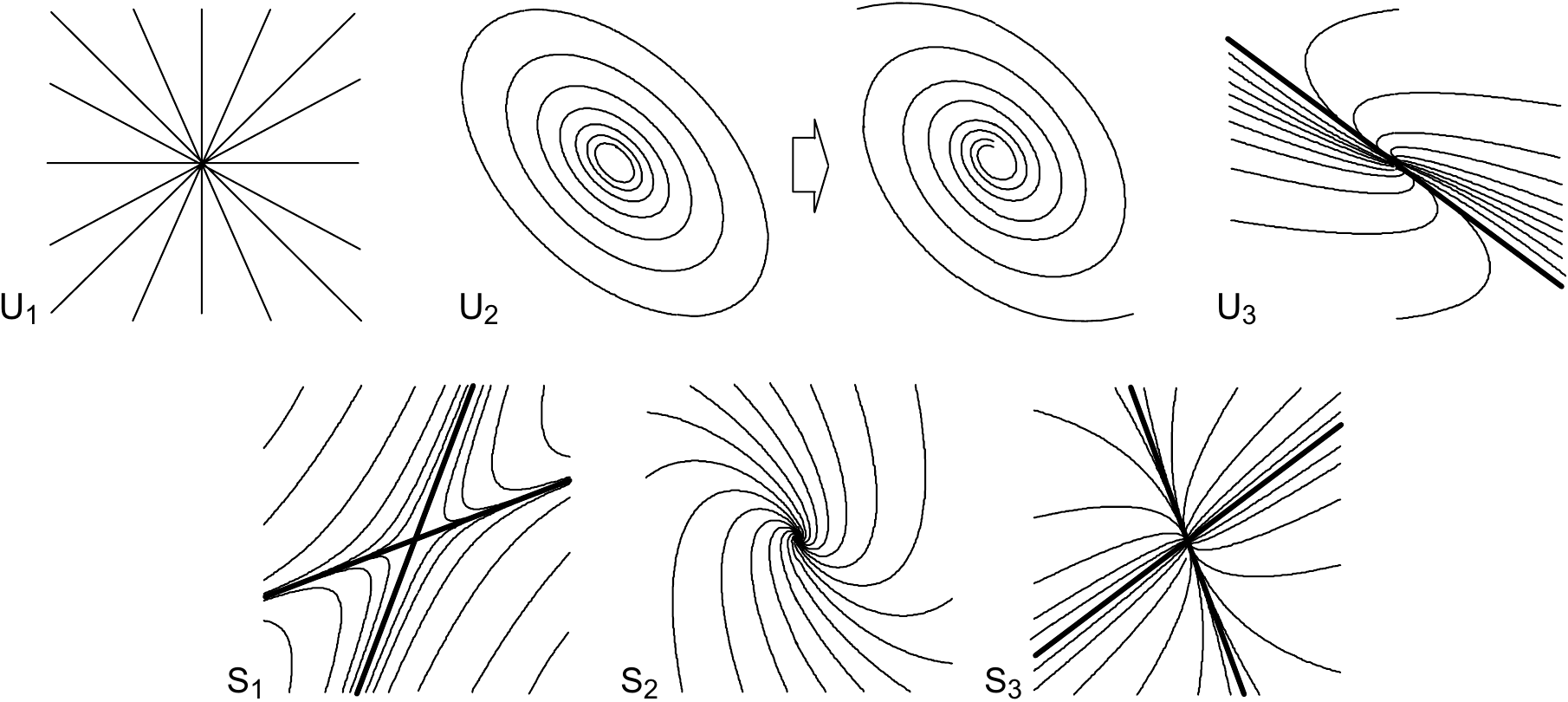}%
\caption{Solutions of Eq. (\ref{dydx}). \ U$_{1}-$U$_{3}$ (S$_{1}-$S$_{3}$)
are unstable (stable) solutions. \ The parameters used in plotting these
solutions are for the unstable solutions: U$_{1}$, $F_{xx}=1,F_{xy}%
=F_{yx}=0,F_{yy}=1$; U$_{2}$, $F_{xx}=1/2-\delta\;(\delta=0\Rightarrow
\delta=0.1),F_{xy}=1,F_{yx}=-1,F_{yy}=-1/2$; U$_{3}$, $F_{xx}=1,F_{xy}%
=1,F_{yx}=-1/4,F_{yy}=0$. \ For the stable solutions the parameters are:
S$_{1}$, $F_{xx}=1,F_{xy}=-1,F_{yx}=1,F_{yy}=-2$; S$_{2}$, $F_{xx}%
=1,F_{xy}=-2,F_{yx}=3,F_{yy}=4$; \ S$_{3}$, $F_{xx}=3,F_{xy}=1/2,F_{yx}%
=1,F_{yy}=2$. \ These parameters were chosen to yield plots that approximately
match the streamlines in Fig. \ref{Fig8}: S$_{1}$ approximately corresponds to
the pattern in Fig. \ref{Fig8}b$_{1}$; S$_{2}$ to the pattern in Fig.
\ref{Fig8}b$_{2}$; S$_{3}$ to the pattern in Fig. \ \ref{Fig8}b$_{3}$. }%
\label{Fig9}%
\end{figure}

The six solutions of Eq. (\ref{dydx}) are illustrated in Fig. \ref{Fig9},
where the three unstable (stable) solutions are labelled U$_{1}$, U$_{2}$, and
U$_{3}$ (S$_{1}$, S$_{2}$, and S$_{3}$).

U$_{1}$ corresponds to $dy/dx=y/x$, and requires $F_{xx}=F_{yy}$ \emph{and}
$F_{xy}=F_{yx}=0$. \ For these values of the parameters $\Delta=0$, and for
this special case $\Lambda=\infty$ rather than $\Lambda=2$. \ When the
stringent conditions on $F$ are no longer met, $\Delta\neq0$, and U$_{1}$
transforms into the stable configuration S$_{2}$ (S$_{3}$), if $\Delta<0$
($\Delta>0$), in both cases preserving index $I$, as expected.

U$_{2}$ corresponds to the condition $F_{yy}=-F_{xx}$. \ When this special
condition is met the streamlines form conic sections: ellipses if $\Delta<0$;
hyperbolas if $\Delta>0$. \ These hyperbolas are organized into the contour
pattern of a saddle point. \ When the special condition on $F$ is no longer
met the ellipses open into the stable spiral S$_{2}$, whereas the
\emph{pattern} of the hyperbolas is maintained as the stable solution S$_{1}$,
although the curves themselves are no longer hyperbolas. \ As expected, in
both cases index $I$ is conserved during the transformation.

U$_{3}$ corresponds to the special condition $F_{xy}F_{yx}=-\frac{1}{4}%
(F_{xx}-F_{yy})^{2}$. \ When this special condition is met $\Delta=0$ and
$\Lambda=2$. \ When the condition on the $F^{\prime}s$ is no longer met
U$_{3}$ transforms into the stable solution S$_{2}$ (S$_{3}$) if $\Delta<0$
($\Delta>0$); again, index $I$ is conserved during the transformation.

Worth noting is that when the special conditions on $F$ are very nearly
satisfied, as can and does happen in a random (or other) field, the pattern of
streamlines can closely approximate the unstable solutions, possibly resulting
in the misleading impression that these solutions have been observed, and are
therefore stable.

For the three stable solutions S$_{1}$, S$_{2}$, and S$_{3}$, indices $I$ and
$\Lambda$ are the following: for S$_{1}$, $I=-1$ and $\Lambda=4$; for S$_{2}$,
$I=+1$ and $\Lambda=0$; for S$_{3}$, $I=+1$ and $\Lambda=4$. \ These index
combinations uniquely determine the streamline pattern, and parallel in some
sense the three index combinations for the three types of stable C points,
\textquotedblleft star\textquotedblright, \textquotedblleft
lemon\textquotedblright, and \textquotedblleft monstar\textquotedblright%
\ [$4,11-13,16$], although the patterns themselves differ for the two cases.
\ The C point combinations are: for the star, $I=-1$ and $\Lambda=3$; for the
lemon, $I=+1$ and $\Lambda=1$; for the monstar, $I=+1$ and $\Lambda=3$.

When $\Lambda=4$, the straight-line streamlines passing through the origin (L
point) are given by%

\begin{equation}
y_{\pm}=\frac{F_{yy}-F_{xx}\pm\sqrt{\Delta}}{2F_{xy}}x\text{.}
\label{Yplusminus}%
\end{equation}

These lines are separatixes that divide the field into four quadrants in which
the sign of the curvature of the streamlines changes across a separatix.

We find that our simulations (Fig. 8) are in full agreement with all of the
above, and that all three stable solutions S$_{1}$, S$_{2}$, and S$_{3}$, are
present for \emph{each} axis $\boldsymbol{\alpha}$, $\boldsymbol{\beta}$, and
$\boldsymbol{\gamma}$, and for \emph{each} type of structure $-$ $\alpha
$-cones, r-rings, and M\"{o}bius strips. \ This reflects the fact that for
each of these structures all three index combinations, $I=-1,\Lambda=4$
(S$_{1}$), $I=1,\Lambda=0$ (S$_{2}$), and $I=1,\Lambda=4$ (S$_{3}$), appear.

\vspace*{-0.3in}

\subsection{Index Summary}

We briefly summarize here the major properties of the $15$ indices that
characterize the M\"{o}bius strips, r-rings, and $\alpha$-cones that surround
L lines.

\paragraph{$I_{\alpha}$, $I_{\beta,\gamma}$, $\Lambda_{\alpha}$,
$\Lambda_{\beta}$, $\Lambda_{\gamma}$.}

These five indices characterize the projections of axes $\boldsymbol{\alpha}$,
$\boldsymbol{\beta}$, and $\boldsymbol{\gamma}$ onto the principal plane
$\Sigma_{0}$. \ Index $I$ measures the winding number of these projections,
$\Lambda$ counts the number of straight streamlines (separatixes) that
terminate (or originate) on the L point. \ For all three axes $I$ takes the
values $\pm1$ and $\Lambda$ takes the values $0,4$. \ If $I=-1$,
$\Lambda\equiv4$, whereas for $I=+1$, $\Lambda=0,4$. \ Each of the three
possible combinations of $I$ and $\Lambda$ defines a unique streamline
pattern, Figs. \ref{Fig8} and \ref{Fig9}.

\paragraph{$d\tau_{\alpha}$, $\tau_{\beta}$, $d\tau_{\beta}$, $\tau_{\gamma}$,
$d\tau_{\gamma}$.}

These five indices characterize the projections of axes $\boldsymbol{\alpha}$,
$\boldsymbol{\beta}$, and $\boldsymbol{\gamma}$ onto the plane $\tau_{0}$,
Fig. \ref{Fig3}. \ $\tau$ measures the winding number of the axes about the
circle of ellipse centers, $\sigma_{0}$. \ For M\"{o}bius strips $\tau=\pm2$,
where $+2$ ($-2$) implies a left-handed (right-handed) two-turn helix, Figs.
\ref{Fig1} and \ref{Fig4}. \ For r-rings (and $\alpha$-cones) $\tau=0$.

$d\tau$ measures the winding number of the tangent to the endpoint curve. \ If
this is a simple closed curve $d\tau=+2$ ($d\tau=-2$), Fig. \ref{Fig4} and
\ref{Fig6}. \ If the endpoint curve generates a figure of eight, which can
occur for axes $\boldsymbol{\beta}$ and $\boldsymbol{\gamma}$, but not for
axis $\boldsymbol{\alpha}$, $d\tau=0$, Fig. \ref{Fig5}. \ When $\tau=\pm2$,
$d\tau=\tau$.

\paragraph{$d\pi_{\alpha}$, $\pi_{\beta}$, $d\pi_{\beta}$, $\pi_{\gamma}$,
$d\pi_{\gamma}$.}

These five indices characterize the projections of axes $\boldsymbol{\alpha}$,
$\boldsymbol{\beta}$, and $\boldsymbol{\gamma}$ onto the plane $\pi_{0}$, Fig.
\ref{Fig3}. \ If the endpoint curve in $\pi_{0}$ encloses the origin (circle
of ellipse centers), which can occur for M\"{o}bius strips and r-rings, but
not for $\alpha$-cones, $\pi=\pm2$, otherwise $\pi=0$. \ $d\pi$ measures the
rotation of the tangent to the $\pi_{0}$ endpoint curve. \ The possible values
of this index for the different axes are the same as those for $d\tau$. \ When
$\pi=\pm2$, $d\pi=\pi$. \ $N_{Z}$, the number of times the endpoint curve in
$\pi_{0}$ crosses the $Z$-axis, is equal to $\Lambda$, the line classification
index, so $N_{Z}$ could replace $\Lambda$; no index or combination of indices
in $\pi_{0}$ and/or $\tau_{0}$, however, can replace $I$.

For a given structure $-$ M\"{o}bius strip, r-ring, or $\alpha$-cone $-$
indices $\tau$ and $d\tau$ of $\tau_{0}$, and indices $\pi$ and $d\pi$ of
$\pi_{0}$, can be, and often are, different, and to a large extent these
indices are also independent of the indices $I$ and $\Lambda$ that describe
the projection onto $\Sigma_{0}$. \ There are, however, index combinations
that do not occur; in the following section these missing combinations are
summarized in the form of selection rules.

\section{STATISTICS}

We discuss here the probabilities of various combinations of the $15$ indices
that characterize the M\"{o}bius strips, r-rings, and $\alpha$-cones that
surround L lines, starting with those combinations for which the probability
is zero. \ These zero probability combinations are summarized below in the
form of binary and ternary selection rules. \ Quaternary and higher-order
rules may also exist. \ Discussed below are simple, systematic methods for
identifying all true binary and ternary rules; in contrast, there do not
appear to be equivalent methods for identifying \emph{true} higher-order rules
(i.e. rules that do not incorporate combinations of lower-order rules).

\subsection{Selection Rules}

The binary and ternary selection rules listed below were either derived from
results discussed earlier, or deduced from the absence of various
configurations in our $10^{6}$ entry database. \ These latter,
\emph{empirical} rules await confirmation from theory.

\begin{figure}
[ph]
\includegraphics[width=0.65\textwidth]%
{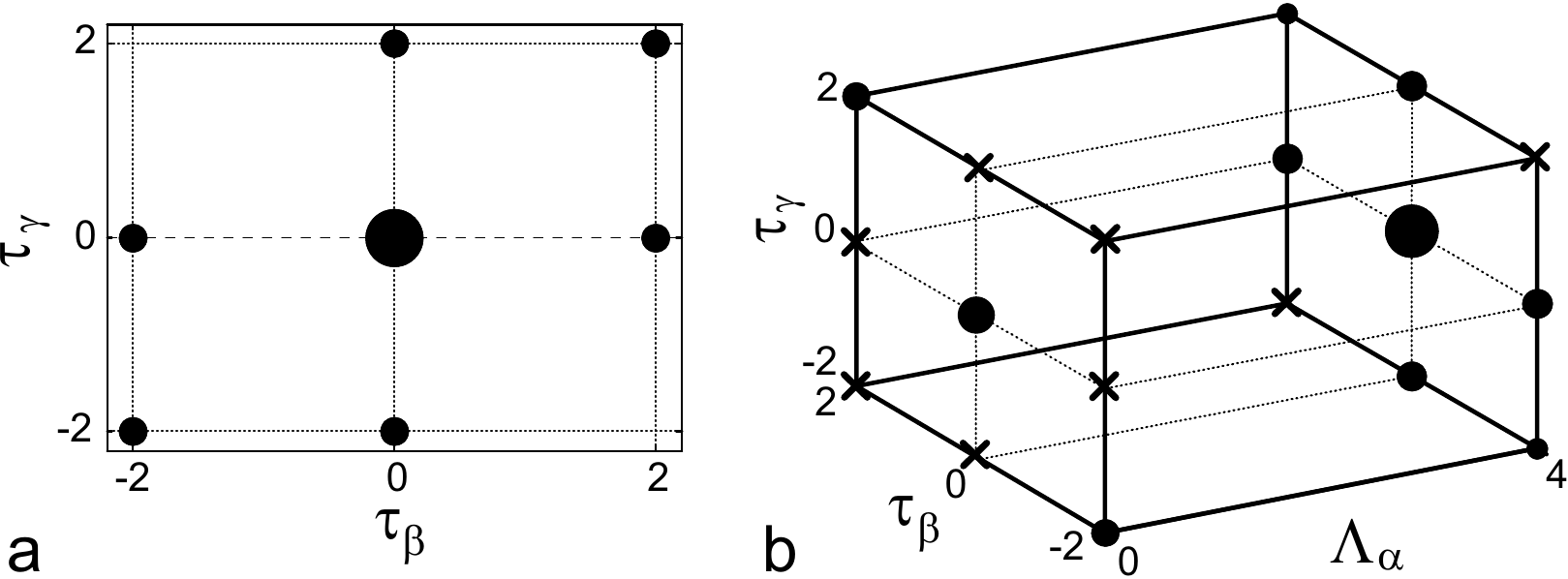}%
\caption{Index correlation plots. \ The relative areas of the circles
approximate the relative probabilities of the different configurations. \ (a)
2D plot of the index combination $\tau_{\beta}-\tau_{\gamma}$. \ Empty grid
points correspond to missing configurations. \ (b) 3D plot of the index
combination $\Lambda_{\alpha}-\tau_{\beta}-\tau_{\gamma}$. \ Missing
configurations are marked by ${\Large \times}$'s. \ The reader may enjoy
identifying the selection rules obtained from these plots.}%
\label{Fig10}%
\end{figure}

Finding the empirical rules was facilitated by the use of 2D and 3D
correlation (scatter) plots. \ Of the $105$ 2D plots of two index
combinations, $17$ had missing configurations that led to the binary selection
rules listed below. \ The number of 3D, three-index plots was $455$; of these
$198$ had missing configurations, most of which, however, involved
combinations of missing two-index configurations already included in the
binary rules. \ The ternary selection rules listed below were obtained from
the $12$ plots that contained true three-index missing configurations (some
plots yielded more than one rule). Typical examples of 2D and 3D correlation
plots with missing configurations are shown in Fig. \ref{Fig10}.%

The binary (ternary) selections rules acting alone reduce the number of
configurations of all $15$ indices from $839,808$ to $16,344$ ($98,864$); both
sets of rules acting together yield $8,248$ allowed configurations, of these
$5,562$ have been observed in our simulations.

\subsubsection{Binary rules}

These rules involve pairs of axes. \ They state that the following two-index
configurations are forbidden:%
\begin{align}
I_{i}  &  =-1\And\Lambda_{i}=0,\;i=\alpha,\beta,\gamma.\tag{Rule 2.1}\\
\left\vert \pi_{i}\right\vert  &  =2\And\Lambda_{i}=0,\;i=\beta,\gamma
.\tag{Rule 2.2}\\
\left\vert \tau_{i}\right\vert  &  =2\And\Lambda_{j}=0,\;ij=\beta\gamma
,\gamma\beta.\tag{Rule 2.3}\\
d\tau_{i}  &  =0\And\Lambda_{i}=0,\;i=\beta,\gamma.\tag{Rule 2.4}\\
d\pi_{i}  &  =0\And\Lambda_{j}=0,\;ij=\beta\gamma,\gamma\beta.\tag{Rule 2.5}\\
\left\vert \tau_{i}\right\vert  &  =2\And d\tau_{i}\neq\tau_{i},\;i=\beta
,\gamma.\tag{Rule 2.6}\\
\left\vert \pi_{i}\right\vert  &  =2\And d\pi_{i}\neq\pi_{i},\;i=\beta
,\gamma.\tag{Rule 2.7}\\
\left\vert \tau_{\beta}\right\vert  &  =2\And\left\vert \tau_{\gamma
}\right\vert =2\And sign\left(  \tau_{\beta}\right)  \neq sign\left(
\tau_{\gamma}\right)  .\tag{Rule 2.8}\\
\left\vert \pi_{\beta}\right\vert  &  =2\And\left\vert \pi_{\gamma}\right\vert
=2\And sign\left(  \pi_{\beta}\right)  \neq sign\left(  \pi_{\gamma}\right)  .
\tag{Rule 2.9}%
\end{align}

\bigskip

The above rules either follow from already discussed, or imply new,
geometrical connections between the structures formed by the different axes
for a given set of wavefield parameters. \ Below we briefly discuss these connections.

\emph{Rule 2.1}. \ This rule is proven in Eq. (\ref{negativeI}).

\emph{Rule 2.2}. $\ $This rule follows from an already discussed geometrical
connection: $\left\vert \pi\right\vert =2$ requires the endpoint curve in
$\pi_{0}$ to cross the $Z$-axis four times, Section III.B.2a and Fig.
\ref{Fig4}; at each such crossing the axis projection is radial, Fig.
\ref{Fig3}, and so $\Lambda=4$.

\emph{Rule 2.3}. $\ $This rule follows from a an already discussed geometrical
connection:$\ \left\vert \tau\right\vert =2$ (i.e. a M\"{o}bius strip)
requires the endpoint curve in $\tau_{0}$ to cross the $Z$-axis four times; at
each such crossing the axis projection is tangential, Fig. \ref{Fig3}.
\ Because, as already noted in Section III.A, the projections of axes
$\boldsymbol{\beta}$ and $\boldsymbol{\gamma}$ onto $\Sigma_{0}$ are
orthogonal, where the $\boldsymbol{\beta}$ ($\boldsymbol{\gamma}$) projection
is tangential the $\boldsymbol{\gamma}$ ($\boldsymbol{\beta}$) projection is
radial, and so $\Lambda_{\gamma}=4$ ($\Lambda_{\beta}=4$)$.$

\emph{Rule 2.4}. \ This rule implies a new geometrical connection: if the
endpoint curve in $\tau_{0}$ is a figure of eight, it must cross the $Z$-axis
four times (see the explanation of Rule $2.3$).

\emph{Rule 2.5}. \ This rule implies a new geometrical connection: if the
endpoint curve of axis $\boldsymbol{\beta}$ ($\boldsymbol{\gamma}$) in
$\pi_{0}$ is a figure of eight, the endpoint curve of axis $\boldsymbol{\gamma
}$ ($\boldsymbol{\beta}$) in $\pi_{0}$ must cross the $Z$-axis four times.

\emph{Rules 2.6 and 2.7}. \ These rules follows from already discussed
geometrical connections: if $\left\vert \tau\right\vert =2$ ($\left\vert
\pi\right\vert =2$) the endpoint curve is elliptical in shape, and therefore
$d\tau=\tau$ ($d\pi=\pi$), Fig. \ref{Fig4}.

\emph{Rule 2.8}. \ This rule implies a new geometrical connection: if both
axis $\boldsymbol{\beta}$ and axis $\boldsymbol{\gamma}$ generate M\"{o}bius
strips, both strips must have the same handedness.

\emph{Rule 2.9}. \ This rule implies a new geometrical connection: if the
endpoint curves in $\pi_{0}$ of axis $\boldsymbol{\beta}$ and axis
$\boldsymbol{\gamma}$ are both elliptical in shape, both curves must have the
same sign.

\subsubsection{Ternary Rules}

These rules involve three different axes. \ They state that the following
three-index configurations are forbidden:%

\begin{align}
\Lambda_{\alpha}  &  =0\And I_{\beta,\gamma}=-1\And\tau_{i}=0,\;i=\beta
,\gamma.\tag{Rule 3.1}\\
\Lambda_{\alpha}  &  =0\And I_{\beta,\gamma}=-1\And d\tau_{i}=0,\;i=\beta
,\gamma.\tag{Rule 3.2}\\
\Lambda_{\alpha}  &  =0\And\Lambda_{i}=0\And\left\vert \tau_{i}\right\vert
=2,\;i=\beta,\gamma.\tag{Rule3.3}\\
\Lambda_{\alpha}  &  =0\And\left\vert \tau_{i}\right\vert =2\And\tau
_{j}=0,\;ij=\beta\gamma,\gamma\beta.\tag{Rule 3.4}\\
\Lambda_{\alpha}  &  =0\And\left\vert \tau_{i}\right\vert =2\And d\tau_{j}%
\neq\tau_{i},\;ij=\beta\gamma,\gamma\beta.\tag{Rule 3.5}\\
\Lambda_{\alpha}  &  =0\And\left\vert \tau_{i}\right\vert =2\And d\tau
_{\alpha}\neq\tau_{i},\;i=\beta,\gamma.\tag{Rule 3.6}\\
\left\vert \tau_{\beta}\right\vert  &  =2\And\tau_{\gamma}=\tau_{\beta}\And
d\tau_{\alpha}\neq\tau_{\beta}.\tag{Rule 3.7}\\
\left\vert \pi_{\beta}\right\vert  &  =2\And\pi_{\gamma}=\pi_{\beta}\And
d\pi_{\alpha}\neq\pi_{\beta}.\tag{Rule 3.8}\\
\left\vert \tau_{i}\right\vert  &  =2\And d\tau_{\alpha}=-\tau_{j}%
,\;ij=\beta\gamma,\gamma\beta.\tag{Rule 3.9}\\
\left\vert \pi_{i}\right\vert  &  =2\And d\pi_{\alpha}=-\pi_{j},\;ij=\beta
\gamma,\gamma\beta.\tag{Rule 3.10}\\
\left\vert \tau_{i}\right\vert  &  =2\And d\tau_{\alpha}=-d\tau_{j}%
,\;ij=\beta\gamma,\gamma\beta.\tag{Rule 3.11}\\
\left\vert \pi_{i}\right\vert  &  =2\And d\pi_{\alpha}=-d\pi_{j}%
,\;ij=\beta\gamma,\gamma\beta. \tag{Rule 3.12}%
\end{align}

All the ternary rules involve connections between $\alpha$-cones and Mobius
strips or r-rings. These connections, all of which are new, appear difficult
to fathom in easily understood, simple geometrical terms. \ For example,
Rule\emph{ }3.1 states that if the streamlines of axis $\boldsymbol{\alpha}$
form a spiral, Fig. \ref{Fig9}-S2, and those of axis $\boldsymbol{\beta}$
(axis $\boldsymbol{\gamma}$) form a saddle point, Fig. \ref{Fig9}-S1, then
axis $\boldsymbol{\beta}$ (axis $\boldsymbol{\gamma}$) itself must generate a
M\"{o}bius strip.

\subsection{Probabilities}

\begin{figure}
[ph]
\includegraphics[width=0.6\textwidth]%
{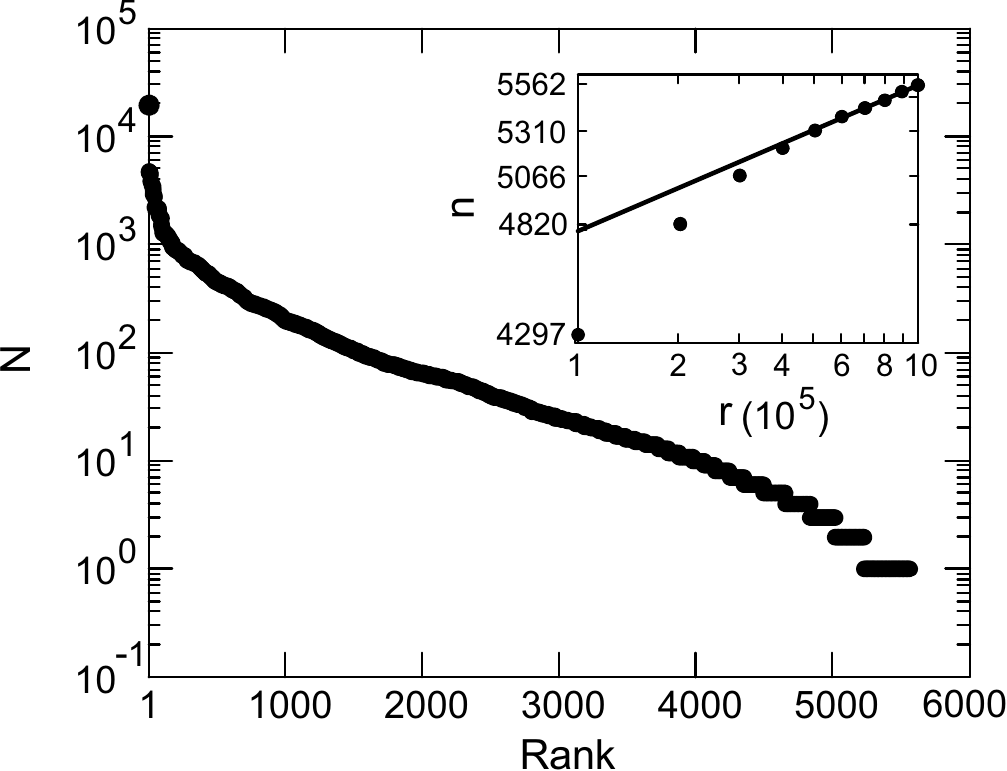}%
\caption{Modified Zipf plot. \ Shown is the number of configurations $N$
observed in our simulation vs. configuration rank $R$. \ The inset shows a
log-log plot of the number of unique configurations, $n$, as a function of the
number of realizations $r$. \ The straight line asymptote is the power law
$n=2,244r^{0.0657}$. \ \ In the four most probable configurations (large dot
at $Rank=1$), which all have substantially the same probability, both axes
$\boldsymbol{\beta}$ and $\boldsymbol{\gamma}$ form M\"{o}bius strips
($\left\vert \tau_{\beta}\right\vert =\left\vert \tau_{\gamma}\right\vert
==2$), the streamlines of axis $\boldsymbol{\alpha}$ form a spiral
($\Lambda_{\alpha}=0$), and those of both axis $\boldsymbol{\beta}$ and axis
$\boldsymbol{\gamma}\ $form saddle points ($I_{\alpha,\beta}=-1,\Lambda
_{\beta}=\Lambda_{\gamma}=4$).}%
\label{Fig11}%
\end{figure}

The relative probabilities of all $15$-index combinations found in our
simulation are displayed in Fig. \ref{Fig11} in the form of a modified Zipf
plot. \ In this plot the number of occurrences of a configuration, $N$, is
plotted vs. the rank of the configuration, $R$, where configurations are
ranked in descending order of probability. \ As can be seen, for $R>1$, $N$
decreases step-wise to one with increasing rank in a quasi-continuous fashion.

But if $N$ decreases quasi-continuously to one, then what meaning can be
attached to the empirical selection rules for which $N=0$? \ The answer is
that there is a qualitative difference between the two-index (three-index) and
the fifteen-index cases, because in the case of the two-index, binary
(three-index, ternary), selection rules, the number of observed occurrences is
either greater than $10^{5}$ ($10^{4}$), or zero.

Shown in the inset of Fig. \ref{Fig11} is the number of unique index
combinations, $n$, as a function of the number of realizations, $r$, where $r$
is incremented sequentially in steps of $10^{5}$. \ As can be seen, $n$
appears to asymptote to the power law $n=2,244r^{0.0657}$. \ The parameters in
this law were chosen such that for $r=10^{6}$, i.e. for our simulation,
$n=5562$, i.e. the number actually observed. \ The binary and ternary
selection rules yield $8,248$ possible configurations. \ Assuming that
higher-order rules do not significantly reduce this number, and that the power
law remains unchanged for much larger $r$, some $4\times10^{8}$ realizations
would be needed to capture all possibilities. \ More realistically, the yield
of new configurations can be expected to fall off with increasing $r$, so that
even $10^{9}$ realizations might not suffice to yield all allowed configurations.%

\begin{figure}
[h]
\includegraphics[width=0.65\textwidth]%
{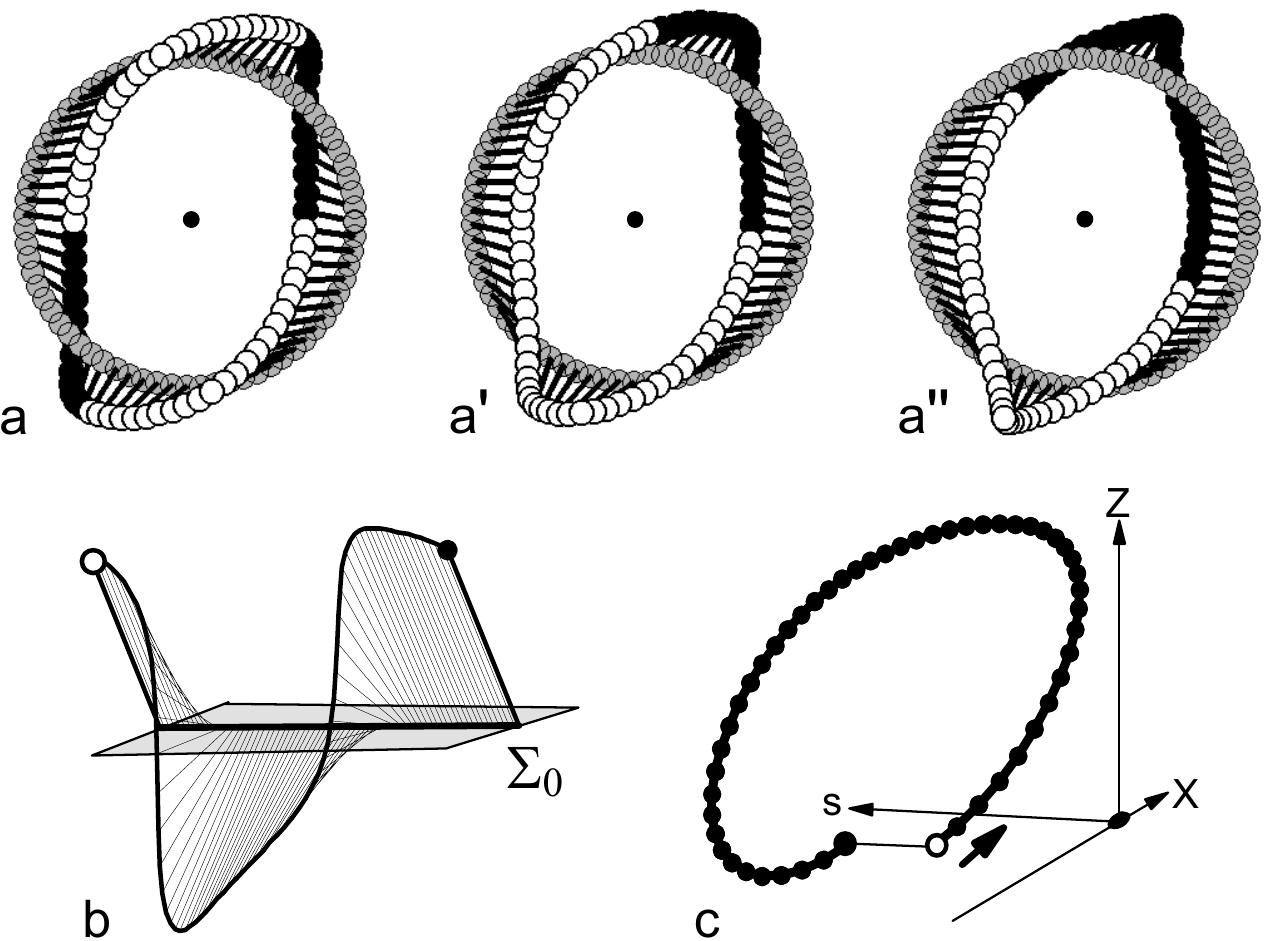}%
\caption{Transformations of M\"{o}bius strips, r-rings, and $\alpha$-cones.
\ As the plane of observation $\Sigma$ is tilted away from $\Sigma_{0}$,
different sets of ellipses occupy the surrounding circle. \ The structures,
indices, and statistics of the M\"{o}bius strips, r-rings, and $\alpha$-cones
generated by these ellipses can change importantly \ with tilt angle. \ For
reasons given below, in the examples shown here the radius $r$ of the
surrounding circle is $r=0.1$, instead of the value used throughout this
report, $r=$ $10^{-4}$; in all the cases shown, when the tilt angle is zero,
the M\"{o}bius strips, r-rings, and $\alpha$-cones are, except for a scale
factor, virtually identical for $r=0.1$ and $r=10^{-4}$. \ (a) The M\"{o}bius
strip in Fig. \ref{Fig1}b for zero tilt angle, i.e. $\Sigma=\Sigma_{0}$. \ As
in Fig. \ref{Fig1}b, axis endpoints are shown by small black (white) circles
if the endpoint lies above (below) the circle of ellipse centers (small gray
circles). \ As can be seen, just like in Fig. \ref{Fig1}b the endpoints and
the ellipse centers form interlocking rings with four crossings, and $\tau
=+2$. \ (a') The strip in (a) for $\Sigma$ tilted away from $\Sigma_{0}$ by
$5^{o}$. \ Here the rings still interlock, but with only two crossings instead
of four, and $\tau=+1$ instead of $\tau=+2$. \ \ (a\textquotedblright) The
strip in (a) for $\Sigma$ tilted away from $\Sigma_{0}$ by $10^{o}$. \ Here
the rings no longer interlock, and $\tau=0$. \ (b) The opened r-ring in Fig.
\ref{Fig1}d' for a tilt angle of $5^{o}$. \ Here the axes oscillate through
only one sinusoidal cycle instead of through two, and $d\tau=+1$ instead of
$+2$ as in Fig. \ref{Fig5}. \ (c) The $\alpha$-cone in Figs. \ref{Fig1}e and
\ref{Fig6} for a $5^{o}$ tilt angle. \ Here the endpoint spiral has only one
turn instead of two, and $d\tau=+1$ instead of $+2$ as in Fig. \ref{Fig6}c.
\ The reason for using $r=0.1$ instead of $r=10^{-4}$ is that the tilt angle
required for a given transformation decreases with $r$, becoming so small for
$r=10^{-4}$ that experimentation becomes impractical. \ For $r_{0}=0.1$,
however, the tilt angle is several degrees, and we use the larger value of $r$
to demonstrate that the transformations described here can be accessible to
experiment. }%
\label{Fig12}%
\end{figure}

Listed below are the probabilities of occurrence in our database of a number
of configurations that may be of special interest. \ These probabilities are
not densities because they do not include the appropriate Jacobian. \ Berry
and Dennis have presented a Jacobian appropriate for some 3D L line properties
[$12$], but this Jacobian does not appear to be suitable for M\"{o}bius
strips, r-rings, and $\alpha$-cones in $\Sigma_{0}$. \ The reason is that
these structures and their statistics can change importantly when the plane of
observation $\Sigma$ is tilted away from $\Sigma_{0}$ by a small but finite,
amount, a property not shared by the Jacobian in [$12$]. \ Because the
probabilities listed below are not densities they are not easily measured
experimentally; they are, however, amenable to calculation. \ Fig. \ref{Fig12}
presents some examples of transformations of M\"{o}bius strips, r-rings, and
$\alpha$-cones under tilt of the plane of observation; these transformed
structures differ importantly from those appearing in $\Sigma_{0}$. \ A more
complete discussion of transformations under tilt will be presented elsewhere.%

From our simulation we find the following:

(\emph{i}) \ As expected, positive and negative values for all indices appear
with equal probabilities.

(\emph{ii}) \ The fraction of L lines or L line segments surrounded by
M\"{o}bius strips is $0.359$, the remaining lines are surrounded by r-rings.
\ This fraction holds for both axis $\boldsymbol{\beta}$ and axis
$\boldsymbol{\gamma}$ strips.

(\emph{iii}) \ The fraction of L lines or L line segments in which \emph{both}
axis $\boldsymbol{\beta}$ and axis $\boldsymbol{\gamma}$ of the surrounding
ellipses generate M\"{o}bius strips (r-rings) is $0.185$ ($0.468$).

(i\emph{v}) \ The fraction of M\"{o}bius strips whose $\pi_{0}$ endpoint
curves form simple closed curves is $0.815$, the remaining endpoint curves are
figures of eight. \ This fraction holds for both axis $\boldsymbol{\beta}$ and
axis $\boldsymbol{\gamma}$.

(\emph{v}) \ The fraction of r-rings whose $\tau_{0}$ or $\pi_{0}$ endpoint
curves form simple closed curves is $0.718$, the remaining endpoint curves
form figures of eight. \ This fraction holds for both axes $\boldsymbol{\beta
}$ and $\boldsymbol{\gamma}$.

(\emph{vi}) \ The fraction of axis $\boldsymbol{\beta}$ Mobius strips
(r-rings) with $I_{\beta,\gamma}=+1$ and $\Lambda_{\beta}=0$ is $0.808$
($0.547$), for the remaining strips (rings) with $I_{\beta,\gamma}=+1$,
$\Lambda_{\beta}=4$. \ The same fractions with subscripts $\beta$ and $\gamma$
interchanged hold for axis $\boldsymbol{\gamma}$ Mobius strips (r-rings).

(\emph{vii}) \ The fraction of Mobius strips (r-rings) with $\Lambda_{\alpha
}=0$ is $0.408$ ($0.228$), for the remaining strips (rings), $\Lambda_{\alpha
}=4$. \ These fractions hold for both axis $\boldsymbol{\beta}$ and axis
$\boldsymbol{\gamma}$ Mobius strips (r-rings).

(\emph{vii}) \ The fraction of $\alpha$-cones with $I_{\alpha}=+1$ and
$\Lambda_{\alpha}=0$ is $0.586$, for the remaining $I_{\alpha}=+1$ cones
$\Lambda_{\alpha}=4$.

Although there are an enormous number of other statistical combinations that
are easily obtained from our database, we believe that the above are likely to
be of the greatest interest since they relate to the major geometric
properties of, and interconnections between, the different structures.

\section{SUMMARY}

The axes of the ellipses surrounding lines of linear polarization, L lines, in
a 3D optical ellipse field have been shown to generate three distinctly
different types of structures. \ In a plane normal to the direction of
polarization of an L point on an L line the major and minor axes of the
surrounding ellipses that lie on small circles surrounding the point generate
M\"{o}bius strips or r-rings (rippled rings), whereas the major axes of these
ellipses generate cones ($\alpha$-cones). \ The M\"{o}bius strips have two
whole twists and can be either right- or left-handed screws. \ These three
different types of structures were characterized by $15$ indices: $12$ winding
numbers that characterize the projections of the different structures onto
three orthogonal planes, $\Sigma_{0}$, $\tau_{0}$, and $\pi_{0}$, and three
indices that characterize the streamlines formed by the projections onto
$\Sigma_{0}$. \ Analytical (numerical) results were given for all indices in
$\Sigma_{0}$ (in $\tau_{0}$ and in $\pi_{0}$). \ Binary and ternary selection
rules were presented that reduce the number of possible configurations from
$839,808$ to $8,248$, of which $5,562$ have been observed in a computer
simulation; probabilities of occurrence were given for the most important of
these configurations.

Experimental measurement of the various structures require determination of
all three orthogonal components of the optical field; such measurements are
currently feasible both in the microwave [$9,10,18,19$], and in the optical
[$20-27$], regions of the spectrum.

\bigskip

\begin{center}
{\large \textbf{References}}
\end{center}

\hspace*{-0.23in}[1] I. Freund, \textquotedblleft Cones, spirals, and Mobius
strips in elliptically polarized light,\textquotedblright\ Opt. Commun.
\textbf{249}, 7$-$22\ (2005).

\hspace*{-0.23in}[2] I. Freund, \textquotedblleft Hidden order in optical
ellipse fields: I. Ordinary ellipses,\textquotedblright\ Opt. Commun.
\textbf{256}, 220$-$241\ (2005).

\hspace*{-0.23in}[3] I. Freund, \textquotedblleft Optical Mobius strips in
three-dimensional ellipse fields: Lines of circular
polarization,\textquotedblright\ Opt. Commun., submitted.

\hspace*{-0.23in}[4] J. F. Nye, \emph{Natural Focusing and Fine Structure of
Light} (IOP Publ., Bristol, 1999).

\hspace*{-0.23in}[5] J. F. Nye, \textquotedblleft Polarization effects in the
diffraction of electromagnetic waves: the role of
disclinations,\textquotedblright\ Proc. Roy. Soc. Lond. A \textbf{387},
105$-$132 (1983).

\hspace*{-0.23in}[6] J. F. Nye, \textquotedblleft Lines of circular
polarization in electromagnetic wave fields,\textquotedblright\ Proc. Roy.
Soc. Lond. A \textbf{389}, 279$-$290 (1983).

\hspace*{-0.23in}[7] J. F. Nye and J. V. Hajnal, \textquotedblleft The wave
structure of monochromatic electromagnetic radiation,\textquotedblright\ Proc.
Roy. Soc. Lond. A \textbf{409}, 21$-$36 (1987).

\hspace*{-0.23in}[8] J. V. Hajnal, \textquotedblleft Singularities in the
transverse fields of electromagnetic waves. I. Theory,\textquotedblright%
\ Proc. Roy. Soc. Lond. A \textbf{414}, 433$-$446 (1987).

\hspace*{-0.23in}[9] J. V. Hajnal, \textquotedblleft Singularities in the
transverse fields of electromagnetic waves. II Observations on the electric
field,\textquotedblright\ Proc. Roy. Soc. Lond. A \textbf{414}, 447$-$468 (1987).

\hspace*{-0.23in}[10] J. V. Hajnal, \textquotedblleft Observations of
singularities in the electric and magnetic fields of freely propagating
microwaves,\textquotedblright\ Proc. Roy. Soc. Lond. A \textbf{430}, 413$-$421 (1990).

\hspace*{-0.23in}[11]. M. V. Berry,\textquotedblleft Geometry of phase and
polarization singularities, illustrated by edge diffraction and the
tides,\textquotedblright\ in Second International Conference on Singular
Optics, M. S. Soskin and M. V. Vasnetsov Eds., Proc. SPIE \textbf{4403,}
1$-$12 (2001).

\hspace*{-0.23in}[12] M. V. Berry and M. R. Dennis, \textquotedblleft
Polarization singularities in isotropic random vector waves,\textquotedblright%
\ Proc. Roy. Soc. Lond. A \textbf{457}, 141$-$155 (2001).

\hspace*{-0.23in}[13] M. V. Berry, \textquotedblleft Index formulae for
singular lines of polarization,\textquotedblright\ J. Opt. A \textbf{6},
675$-$678 (20044.

\hspace*{-0.23in}[14] M. Born and E. W. Wolf, \emph{Principles of Optics}
(Pergamon Press, Oxford, 1959).

\hspace*{-0.23in}[15] I. Freund, \textquotedblleft Coherency matrix
description of optical polarization singularities,' J. Opt. A \textbf{6},
S229$-$S234 (2004).

\hspace*{-0.23in}[16] M. V. Berry and J. H. Hannay, \textquotedblleft Umbilic
points on Gaussian random surfaces,\textquotedblright\ J. Phys. A \textbf{10},
1809$-$1821 (1977).

\hspace*{-0.23in}[17] L. R. Ford, \emph{Differential Equations} (McGraw-Hill,
New York, 1955)

\hspace*{-0.23in}[18] S. Zhang and A. Z. Genack, \textquotedblleft Statistics
of Diffusive and Localized Fields in the Vortex Core,\textquotedblright\ Phys.
Rev. Lett. \textbf{99}, 203901 (2007).

\hspace*{-0.23in}[19] S. Zhang, B. Hu, P. Sebbah, and A. Z. Genack,
\textquotedblleft Speckle Evolution of Diffusive and Localized
Waves,\textquotedblright\ Phys. Rev. Lett. \textbf{99}, 063902 (2007).

\hspace*{-0.23in}[20] R. Dandliker, I. Marki, M. Salt, and A. Nesci,
\textquotedblleft Measuring optical phase singularities at subwavelength
resolution,\textquotedblright\ J. Optics A \textbf{6}, S189$-$S196 (2004).

\hspace*{-0.23in}[21] P. Tortora, R. Dandliker, W. Nakagawa, and L. Vaccaro,
\textquotedblleft Detection of non-paraxial optical fields by optical fiber
tip probes,\textquotedblright\ Opt. Commun. \textbf{259}, 876$-$882 (2006).

\hspace*{-0.23in}[22] C. Rockstuhl, I. Marki, T. Scharf, M. Salt, H. P.
Herzig, and R. Dandliker, \textquotedblleft High resolution interference
microscopy: A tool for probing optical waves in the far-field on a nanometric
length scale,\textquotedblright\ Current Nanoscience \textbf{2}, 337$-$350 (2006).

\hspace*{-0.23in}[23] P. Tortora, E. Descrovi, L. Aeschimann, L. Vaccaro, H.
P. Herzig, and R. Dandliker, \textquotedblleft Selective coupling of HE11 and
TM01 modes into microfabricated fully metal-coated quartz
probes,\textquotedblright\ Ultramicroscopy \textbf{107}, 158$-$165 (2007).

\hspace*{-0.23in}[24] K. G. Lee, H. W. Kihm, J. E. Kihm, W. J. Choi, H. Kim,
C. Ropers, D. J. Park, Y. C. Yoon, S. B. Choi, H. Woo, J. Kim, B. Lee, Q. H.
Park, C. Lienau C, and D. S. Kim, \textquotedblleft Vector field microscopic
imaging of light,\textquotedblright\ Nature Photonics \textbf{1}, 53$-$56 (2007).

\hspace*{-0.23in}[25] Z. H. Kim and S. R. Leone, \textquotedblleft
Polarization-selective mapping of near-field intensity and phase around gold
nanoparticles using apertureless near-field microscopy,\textquotedblright%
\ Opt. Express \textbf{16}, 1733$-$1741 (2008).

\hspace*{-0.23in}[26] M. Burresi, R. J. Engelen, A. Opheij, D. van Oosten, D.
Mori, T. Baba, and L. Kuipers, \textquotedblleft Observation of Polarization
Singularities at the Nanoscale,\textquotedblright\ Phys. Rev. Lett.
\textbf{102}, 033902 (2009).

\hspace*{-0.23in}[27] R. J. Engelen, D. Mori, T. Baba, and L. Kuipers,
\textquotedblleft Subwavelength Structure of the Evanescent Field of an
Optical Bloch Wave,\textquotedblright\ Phys. Rev. Lett. \textbf{102}, 023902
(2009); Erratum: ibid. 049904 (2009).

\end{document}